# Asteroid 16 Psyche: Shape, Features, and Global Map

Michael K. Shepard[1][2], Katherine de Kleer[3], Saverio Cambioni[3], Patrick A. Taylor[4], Anne K. Virkki[4], Edgard G. Rivera-Valentin[4], Carolina Rodriguez Sanchez-Vahamonde[4], Luisa Fernanda Zambrano-Marin[5], Christopher Magri[5], David Dunham[6], John Moore[7], Maria Camarca[3]

## Abstract

We develop a shape model of asteroid 16 Psyche using observations acquired in a wide range of wavelengths: Arecibo S-band delay-Doppler imaging, Atacama Large Millimeter Array (ALMA) plane-of-sky imaging, adaptive optics (AO) images from Keck and the Very Large Telescope (VLT), and a recent stellar occultation. Our shape model has dimensions 278 (−4/+8 km) x 238(−4/+6 km) x 171 km (−1/+5 km), an effective spherical diameter $D_{eff}$ = 222 −1/+4 km, and a spin axis (ecliptic lon, lat) of (36°, -8°) ± 2°. We survey all the features previously reported to exist, tentatively identify several new features, and produce a global map of Psyche. Using 30 calibrated radar echoes, we find Psyche's overall radar albedo to be 0.34 ± 0.08 suggesting that the upper meter of regolith has a significant metal (*i.e.*, Fe-Ni) content. We find four regions of enhanced or complex radar albedo, one of which correlates well with a previously identified feature on Psyche, and all of which appear to correlate with patches of relatively high optical albedo. Based on these findings, we cannot rule out a model of Psyche as a remnant core, but our preferred interpretation is that Psyche is a differentiated world with a regolith composition analogous to enstatite or CH/CB chondrites and peppered with localized regions of high metal concentrations. The most credible formation mechanism for these regions is ferrovolcanism as proposed by Johnson et al. [Nature Astronomy vol 4, January 2020, 41-44.].

---

[1]Corresponding author mshepard@bloomu.edu
[2]Bloomsburg University, 400 E. Second St., Bloomsburg, PA 17815, USA
[3]California Institute of Technology, Pasadena, CA 91125, USA
[4]Arecibo Observatory, University of Central Florida, Arecibo, PR 00612, USA
[5]University of Maine Farmington, Farmington, ME, 04938 USA.
[6]International Occultation Timing Association, Greenbelt, MD 20768-0423, USA

## 1. INTRODUCTION

Asteroid 16 Psyche is the largest Tholen [1984] M-class asteroid and a target of the NASA Discovery mission *Psyche* [Elkins-Tanton et al., 2017; 2020]. Visible and near-infrared spectra [Bell et al. 1989], optical polarimetry [Dollfus et al. 1979], thermal observations [Matter et al. 2013], and radar observations [Ostro et al., 1985, Magri et al. 2007a, Shepard et al., 2008; Shepard et al., 2017] all support the hypothesis that its surface to near-surface is dominated by metal (*i.e.*, iron and nickel). However, recent spectral detections of silicates [Hardersen et al. 2005; Ockert-Bell et al. 2008, 2010; Sanchez et al. 2017] and hydrated mineral phases [Takir et al. 2017] have complicated this interpretation and suggest it to be more complex than previously assumed.

Recent studies of Psyche have converged to size estimates (effective diameter) between 220 and 230 km, and to spin poles within a few degrees of (36°, -8°) (ecliptic lon, lat) (Shepard et al. 2017; Drummond et al. 2018; Viikinkoski et al. 2018; Ferrais et al. 2020). Given this size, mass estimates have led to a consensus that Psyche's overall bulk density is between 3,400 and 4,100 kg m$^{-3}$ [Elkins-Tanton et al. 2020] which places constraints on its internal structure, composition, and possible origin.

Three major hypotheses for the formation and structure of Psyche are currently debated. In simplified terms they are: (1) the early interpretation that Psyche is the stripped remnant core of an ancient planetesimal, dominated by an iron composition [e.g. Bell et al. 1989; Asphaug et al. 2006]; (2) Psyche is a reaccumulated pile of rock and metal, a type of low-iron pallasite or mesosiderite parent body derived from repeated impacts [Davis et al. 1999; Viikinkoski et al. 2018]; and (3) Psyche is either an iron [Abrahams and Nimmo, 2019] or a silicate-iron differentiated body [Johnson et al. 2020] with surface eruptions of metallic iron via ferrovolcanism.

The earliest model for Psyche is that it is a remnant metal core left after a large impact stripped the crust and mantle from an early protoplanet [Bell et al. 1989; Davis et al. 1999]. However, problems with this interpretation have been difficult to overcome. Davis et al. [1999] suggested that an event of this magnitude was unlikely in first 500 Ma of solar system formation and would have left dozens of family fragments that should be detectable but have not been found. The presence of silicates, and especially hydrated phases detected on Psyche has also created doubt. Only recently have new hypervelocity impact experiments on iron bodies suggested a way to systematically explain these signatures [Libourel et al. 2020]. Perhaps the most problematic datum is Psyche's bulk density. Iron has a grain density of ~7,900 kg m$^{-3}$, yet multiple estimates of Psyche's mass yield an overall bulk density of half that value [Elkins-Tanton et al. 2020]. If it is a remnant core, it must also be a rubble pile or extremely porous, and it is not known if this is possible for an object as large as Psyche. Although still possible, it is fair to say that this model for Psyche has fallen out of favor [Elkins-Tanton et al. 2020].

As an alternative, Davis et al. [1999] suggested that, while Psyche was shattered by early impacts, it was not completely disrupted. As a result, the surface should preserve a mesosiderite-like mix of metal and silicate material with a still-intact core. The detections of orthopyroxenes and other possible silicates in the spectra of Psyche [Hardersen et al. 2005; Ockert-Bell et al. 2010] are consistent with this. The detection of spectral hydroxyl features [Takir et al. 2017] is inconsistent with this violent early genesis, but easily explained by more recent impacts with primitive objects, similar to the scenario envisioned for Vesta [Prettyman et al. 2012; Reddy et al. 2012; Shepard et al. 2015] and later demonstrated experimentally by Libourel et al. [2020].

Recently, two groups have proposed that Psyche is an object that has experienced a type of iron volcanism, or ferrovolcanism. Abrahams and Nimmo [2019] propose a mechanism in which molten iron erupts onto the brittle surface of a pure iron-nickel remnant as it cools from the outside inward. They predict that iron volcanos are more likely to be associated with impact craters, or if there is an overlying silicate layer, the iron may be deposited intrusively, as a diapir. Conceivably, these intrusions could later be uncovered by impacts. However, the underlying assumption of this model is that Psyche is largely metallic throughout, possibly covered by a thin layer of silicate impact debris. This seems incompatible with Psyche's overall bulk density.

Johnson et al. [2020] propose that Psyche is a differentiated object with a relatively thin silicate mantle and iron core. As it cooled, volatile-rich molten iron was injected into the overlying mantle and, in favorable circumstances, erupted onto the surface. This model is consistent with Psyche's overall bulk density and the observation of silicates on its surface. They make no specific predictions about the location of eruption centers except that they are more likely where the crust and mantle are thinnest.

In both ferrovolcanic models, the presence of volatiles, *e.g.,* sulfur, is critical for lowering the viscosity of the iron melt and providing the pressure needed to drive the melt onto the surface. In the Johnson et al. model, the thickness of the mantle/crust is also critical; if it is too thick, the melt will never make it to the surface.

In this paper, we refine the size and shape of Psyche using previously published and newly obtained data sets. We examine topographical and optical albedo features discussed in previous work [Shepard et al. 2017; Viikinkoski et al. 2018; Ferrais et al. 2020], produce a global map, and overlay it with radar albedo values, a proxy for metal concentrations in the upper meter of the regolith. We find correlations between centers of high radar and optical albedos and discuss the consequences for these models of Psyche's formation and structure.

## 2. DATA SETS AND SHAPE MODEL

In this section, we briefly describe the datasets used in our shape model, our methods, and final model results. Dataset details are listed in **Tables 1-5. Figures 1-6** illustrate both the image datasets and subsequent model fits to them.

### 2.1. Arecibo S-band Delay-Doppler Imaging and Calibrated Echoes

For shape modeling, we use 18 delay-Doppler images acquired by the Arecibo S-band (2380 MHz or 12.6 cm wavelength) radar in 2015 [Shepard et al. 2017]. These images have a spatial scale (in range) of 7.5 km/pixel, sample frequency of 50 Hz per pixel (Doppler), and were centered in the southern mid-latitudes of Psyche. Details of those observations are in **Table 1** and illustrated in **Figure 1**.

Later, we utilize 30 calibrated continuous wave (CW) echoes to examine the radar reflectivity of the upper meter of Psyche's regolith. These include 16 observations in 2005 [Shepard et al. 2008], five in 2015 [Shepard et al. 2017] and 9 new observations in 2017. The 2005 and 2015 observations were centered in the southern mid-latitudes, and the 2017 observations were the first radar observations to probe the northern mid-latitudes. Details of the observations are in **Table 2.**

**Table 1. Psyche Delay-Doppler Imaging**

| Run | Date & Time | Lat,Lon(°) |
|---|---|---|
| 1 | 2015 Nov 29 04:37 | -47, 301 |
| 2 | 2015 Nov 29 06:20 | -47, 154 |
| 3 | 2015 Nov 30 04:32 | -47, 48 |
| 4 | 2015 Nov 30 06:16 | -47, 260 |
| 5 | 2015 Dec 01 04:58 | -47, 113 |
| 6 | 2015 Dec 01 05:54 | -47, 32 |
| 7 | 2015 Dec 02 04:33 | -47, 249 |
| 8 | 2015 Dec 02 05:46 | -47, 145 |
| 9 | 2015 Dec 03 05:49 | -48, 241 |
| 10 | 2015 Dec 04 04:12 | -48, 121 |
| 11 | 2015 Dec 04 05:56 | -48, 333 |
| 12 | 2015 Dec 05 04:18 | -48, 213 |
| 13 | 2015 Dec 05 05:50 | -48, 82 |
| 14 | 2015 Dec 06 04:04 | -48, 334 |
| 15 | 2015 Dec 06 06:00 | -48, 179 |
| 16 | 2015 Dec 07 03:58 | -48, 84 |
| 17 | 2015 Dec 07 05:40 | -48, 298 |
| 18 | 2015 Dec 09 05:14 | -49, 178 |

Earth-Psyche range was 1.70 AU. All dates and times refer to the UTC midpoint of the receive time. Each delay-Doppler image incorporates 28 minutes of echo integration. Lat/Lon refers to the sub-radar body-centered latitude and longitude at the time noted (light-time corrected).

**Figure 1.** Psyche delay-Doppler imaging of 2015 and shape model fits. Images are grouped in column triads: at left is the image data, center is the simulated image data using the model shape and aspect shown in the plane-of-sky view on the right. The images are ordered in rotation sequence, left to right, top to bottom. Run numbers are indicated and can be matched to Table 1 for dates, times, and sub-radar body-centered longitudes. Doppler frequency is along the horizontal axis of the delay-Doppler images (0 in center, positive on the left, ±1000 Hz per image, 50 Hz/pixel), delay increases from top to bottom (307.5 km total per image, 7.5 km/pixel). The spin axis is indicated by the long arrow, the long-axis (lon 0°) is marked by the short (red) peg and the intermediate axis (lon 90°) is indicated by the longer (green) peg. Small white arrows on the images point to the evidence for a large crater (Eros).

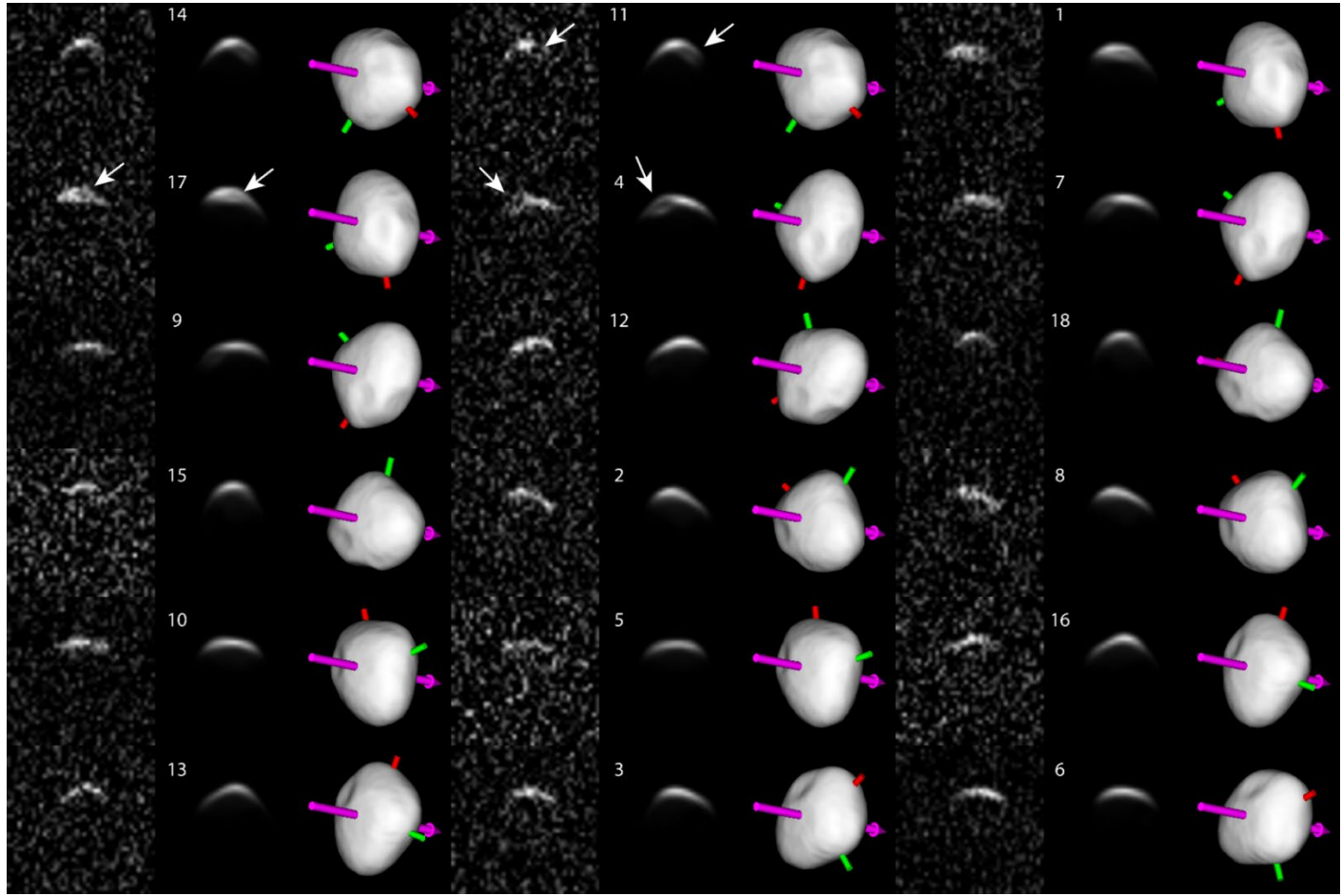

**Table 2. Psyche Echo Power (CW) Radar Observations**

| Epoch (UT) | SNR | Lat, Lon (°) | $\sigma_{OC}$ (km$^2$) | $\mu_c$ | Area (km$^2$) | $\hat{\sigma}_{OC}$ |
|---|---|---|---|---|---|---|
| **2017 Feb 28 04:48** | **19** | **+37, 302** | **20,300** | **0.18** | **41,370** | **0.49** |
| 2017 Mar 01 04:44 | 10 | +37, 49 | 11,400 | 0.0 | 41,570 | 0.27 |
| 2017 Mar 02 04:48 | 13 | +37, 144 | 12,100 | 0.0 | 41,190 | 0.29 |
| 2017 Mar 03 05:30 | 6 | +37, 186 | 9,000 | 0.0 | 40,040 | 0.22 |
| ***2017 Mar 05 04:57*** | 14 | +36, 75 | 14,900 | 0.0 | 41,910 | 0.36 |
| 2017 Mar 06 04:51 | 8 | +36, 184 | 10,400 | 0.2 | 39,830 | 0.26 |
| 2017 Mar 07 05:09 | 9 | +36, 256 | 12,000 | 0.0 | 42,840 | 0.28 |
| 2017 Mar 08 05:05 | 9 | +36, 6 | 9,600 | 0.1 | 39,290 | 0.29 |
| 2017 Mar 11 04:00 | 11 | +35, 41 | 11,700 | 0.2 | 40,690 | 0.29 |
| 2015 Nov 29 05:28 | 24 | -47, 228 | 13,200 | 0.09 | 43920 | 0.30 |
| 2015 Dec 03 04:42 | 27 | -48, 337 | 12,800 | 0.03 | 44040 | 0.29 |
| 2015 Dec 04 05:04 | 12 | -48, 41 | 11,200 | 0.1 | 44230 | 0.28 |
| ***2015 Dec 06 04:54*** | 22 | -48, 262 | 16,000 | 0.21 | 44960 | 0.36 |
| **2015 Dec 07 04:48** | **33** | **-48, 12** | **18400** | **0.05** | **43710** | **0.42** |
| 2005 Nov 12 05:51 | 15 | -45, 181 | 12,200 | 0.18 | 41960 | 0.29 |
| 2005 Nov 12 06:05 | 15 | -45, 161 | 13,200 | 0.18 | 42150 | 0.31 |
| 2005 Nov 13 05:37 | 17 | -45, 302 | 15,900 | 0.04 | 44140 | 0.36 |
| ***2005 Nov 13 05:51*** | 16 | -45, 282 | 15,400 | 0.02 | 44420 | 0.35 |
| 2005 Nov 13 06:59 | 14 | -45, 184 | 10,900 | 0.00 | 42020 | 0.26 |
| 2005 Nov 13 07:14 | 13 | -45, 163 | 12,200 | 0.1 | 42080 | 0.29 |
| **2005 Nov 14 05:44** | **22** | **-45, 32** | **21,600** | **0.00** | **43330** | **0.50** |
| **2005 Nov 14 05:58** | **20** | **-45, 12** | **21,000** | **0.18** | **42860** | **0.49** |
| ***2005 Nov 14 07:08*** | 14 | -45, 273 | 16,000 | 0.16 | 44460 | 0.36 |
| 2005 Nov 14 07:22 | 11 | -45, 253 | 9,500 | 0.4 | 44120 | 0.22 |
| **2005 Nov 15 06:50** | **21** | **-46, 38** | **21,300** | **0.00** | **43600** | **0.49** |
| **2005 Nov 15 07:05** | **24** | **-46, 18** | **22,500** | **0.00** | **43020** | **0.52** |
| ***2005 Nov 16 05:26*** | 17 | -46, 259 | 14,000 | 0.00 | 44360 | 0.32 |
| 2005 Nov 16 05:40 | 18 | -46, 240 | 14,200 | 0.03 | 43970 | 0.32 |
| 2005 Nov 16 06:46 | 18 | -46, 146 | 14,400 | 0.05 | 42980 | 0.34 |
| **2005 Nov 16 07:00** | **20** | **-46, 126** | **18,000** | **0.07** | **43750** | **0.42** |

Earth-Psyche range was 1.73, 1.70, 2.24 AU for 2005, 2015, and 2017 runs, respectively. Times are UTC midpoints of each run's receive cycle. Signal-to-noise (SNR) is shown for matched filters. Lat/Lon refers to the sub-radar body-centered latitude and longitude at the time noted (light-time corrected). **σ₀c** is the OC radar cross-section; uncertainties are conservatively estimated to be ±25% based on historic calibration and pointing uncertainties. μ$_c$ is the radar polarization ratio (±0.1 for values with 1 significant figure and ±0.02 for the others). Area is the projected area of the shape model visible at the time of each run (±3%). $\hat{\sigma}_{OC}$ is the OC radar albedo. **Dates/times in bold have high radar albedos.** ***Dates/times in bold italics indicate echoes that appear bifurcated.***

**Figure 2.** Continuous wave observations of Psyche in 2017. The x-axis is Doppler frequency, the y-axis is echo power in standard deviations. The data have been smoothed to an effective resolution of 100 Hz. The solid line is the OC radar echo; the dashed line is the SC echo. Each run integrates 37 minutes of echo reception. Times, sub-radar locations, echo powers, and polarization ratios of each observation are given in Table 2. The March 5 run shows a possible bifurcated echo.

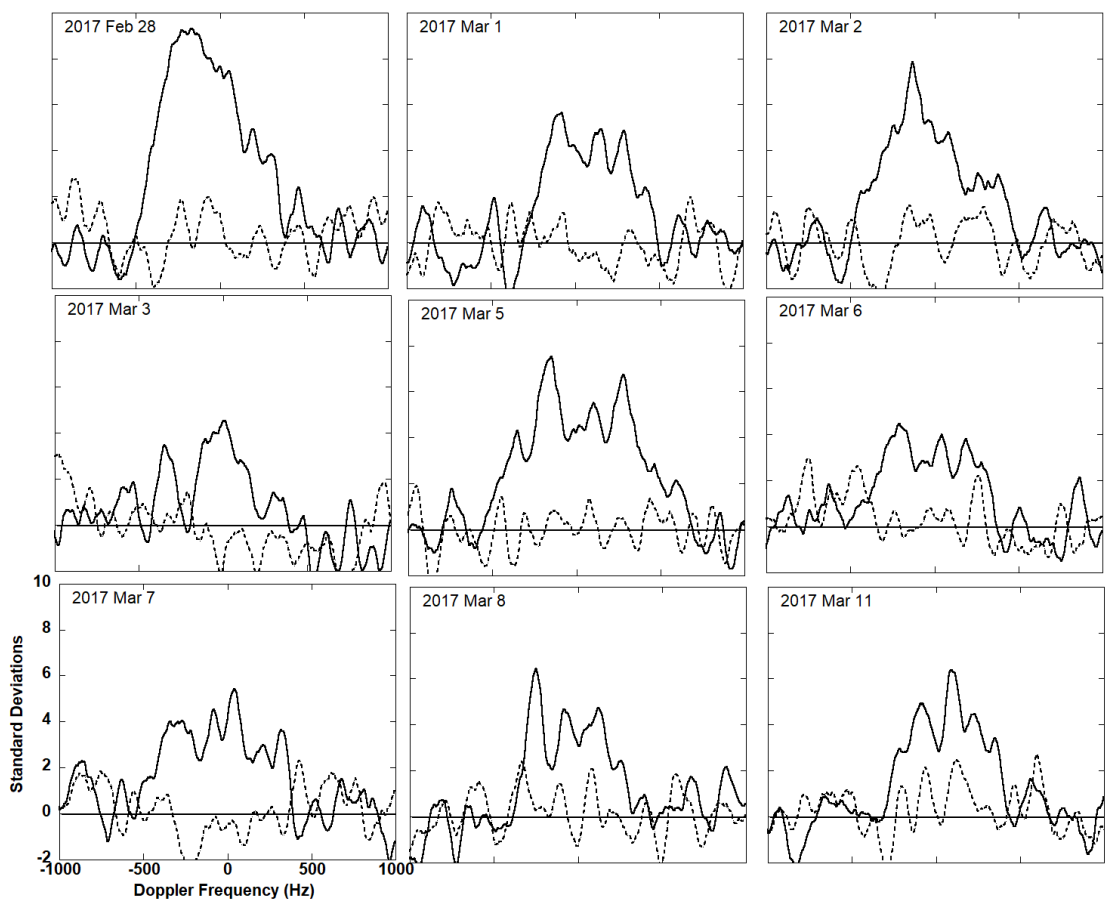

*2.2 Atacama Large Millimeter Array*

For the first time ever, Psyche was imaged with the Atacama Large (sub-) Millimeter Array (ALMA) in the Atacama Desert in Chile on UT June 19, 2019 over ~2/3 of its rotation at a (model) sub-observer latitude of -14 degrees (de Kleer et al., 2021). Spatially-resolved thermal emission data were obtained in ALMA's Band 6 at a central frequency of 232 GHz (1.3 mm). ALMA was in an extended configuration at the time of observation with a maximum baseline of 16.2 km yielding a spatial resolution of 0.02" or 30 km at Psyche. The data were reduced and calibrated using the ALMA pipeline. Additional calibration and imaging of the data were performed using the Common Astronomy Software Applications (CASA) package (McMullin et al. 2007), employing a custom iterative imaging and calibration routine. A more extensive description of the observations as well as calibrations and imaging methods is given in de Kleer et al. (2021). In this paper, we use the observations primarily to constrain Psyche's overall shape and size. The observing geometry is summarized here in Table 3 and the image data are shown in Figure 3.

### 2.3. Adaptive Optics Images

We use 10 observations of Psyche acquired by the Very Large Telescope (VLT) Spectro-Polarimetric High-contrast Exoplanet Research (SPHERE) instrument in 2018 and 2019 [Viinkinkoski et al. 2018; Ferrais et al. 2020]. The observations were taken in the N_R filter centered on 0.646 μm. The 2018 observations were at a near-polar northern aspect and the 2019 observations were at a near-equatorial aspect. Details are listed in **Table 4** and illustrated in **Figure 4.**

For our fit, we also included four adaptive optics (AO) images acquired in 2015 [Shepard et al. 2017; Drummond et al. 2018]. These images (Kp band, 2.1 μm) were taken at the Keck II telescope at roughly the same time and aspect (southern midlatitudes) as the 2015 Arecibo radar imaging runs. After the model was completed, we acquired new Keck observations in December 2020 at nearly identical aspects. These were used only to check the model. Details are listed in **Table 4** and illustrated in **Figure 4.**

**Table 3. ALMA Observations**

**2019 Jun 19**

| Image | Time | Lon($^o$) |
|:-----:|:----:|:---------:|
| 1 | 6:35 | 200 |
| 2 | 6:40 | 193 |
| 3 | 6:46 | 185 |
| 4 | 6:52 | 176 |
| 5 | 6:58 | 168 |
| 6 | 7:04 | 160 |
| 7 | 7:19 | 137 |
| 8 | 7:24 | 131 |
| 9 | 7:29 | 124 |
| 10 | 7:34 | 117 |
| 11 | 7:38 | 111 |
| 12 | 8:05 | 72 |
| 13 | 8:11 | 64 |
| 14 | 8:17 | 55 |
| 15 | 8:23 | 47 |
| 16 | 8:28 | 38 |
| 17 | 8:34 | 31 |
| 18 | 8:49 | 9 |
| 19 | 8:54 | 2 |
| 20 | 9:00 | 353 |
| 21 | 9:05 | 346 |
| 22 | 9:09 | 340 |

Data from de Kleer et al. [2021]. A total of 88 scans, each 55 seconds or less were made. The images listed here and shown in Figure 3 are sums of 3-5 adjacent scans, giving a total integration time for each image of 3-6 minutes with a SNR of 40-50 after processing. Times are 2019 Jun 19 UTC midpoints of each image sum. Lon refers to the model sub-observer longitude (latitude -14$^o$) at the time of the observation (light-time corrected). The interferometric beam was essentially circular and roughly 0.02 arcseconds in half-power beam width, corresponding to a diameter of 30 km at Psyche, 2.04 AU from Earth.

**Figure 3.** Psyche ALMA image data in 2019 ordered in rotation sequence, left-to-right, top-to-bottom. Images are in column pairs: the image data is on the left and the simulated model view is on its immediate right. Sub-observer latitude was -14°; longitudes are noted above the model on select images. Dates, times, resolutions, and aspect information is in Table 3. The approximate beam size (30 km) is indicated and the image scale is 4.4 km per pixel. The cross-hatched pattern visible in the data is an artifact of the imaging process. Arrows with letters indicate the position of features mentioned in the text and Table 6 (first letters only are shown). One feature (Xray) is labeled on ALMA images, but is not seen in the model.

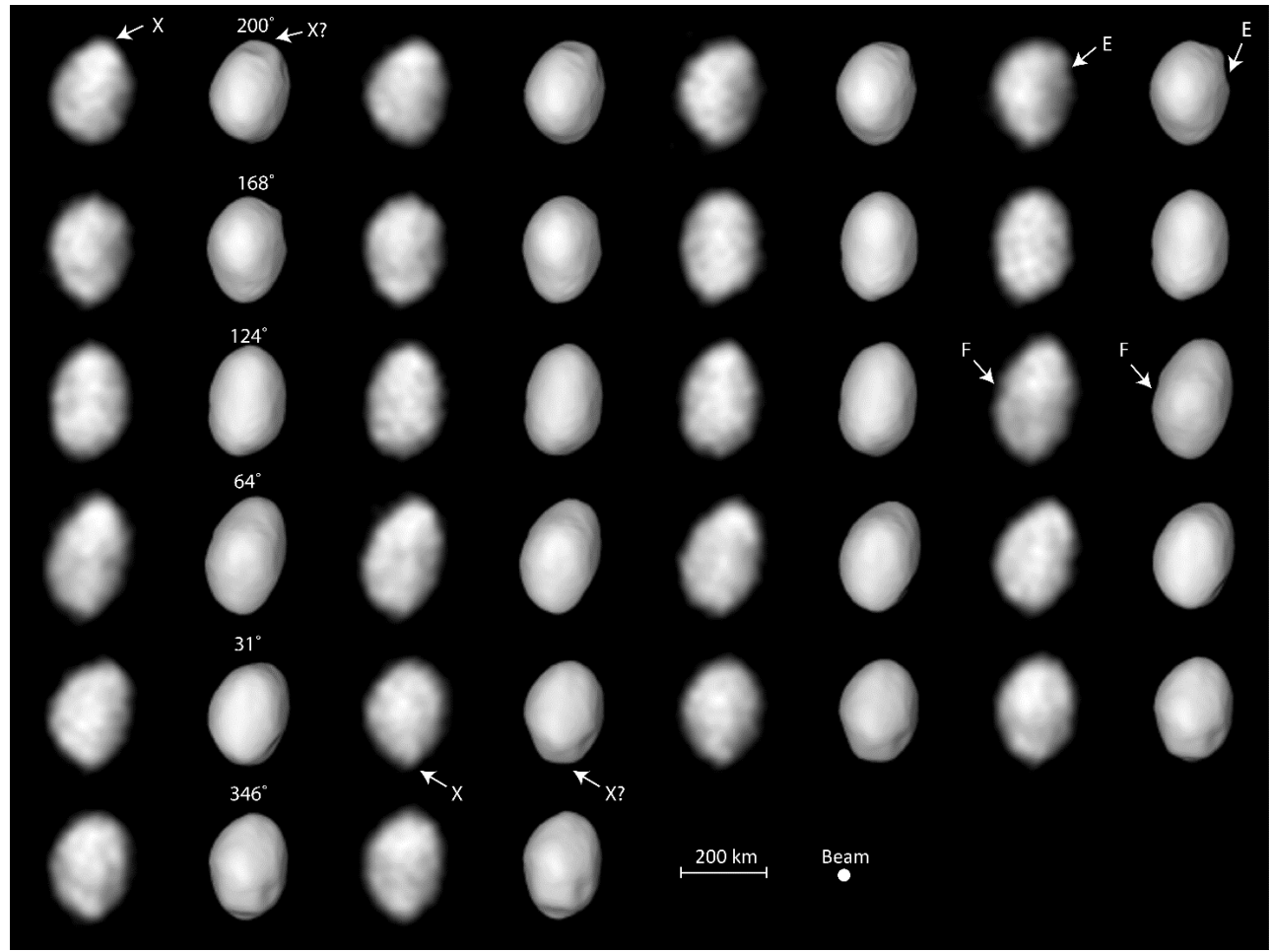

**Table 4. Adaptive Optics Images**

| Date & Time | Source | Band | Lat,Lon(°) |
|---|---|---|---|
| 2015 Dec 25 08:54 | Keck | Kp | -52, 38 |
| 2015 Dec 25 09:50 | Keck | Kp | -52, 318 |
| 2015 Dec 25 10:35 | Keck | Kp | -52, 254 |
| 2015 Dec 25 11:23 | Keck | Kp | -52, 186 |
| 2020 Dec 05 12:09 | Keck | J | -49, 231 |
| 2020 Dec 05 12:14 | Keck | H | -49, 223 |
| 2020 Dec 05 12:19 | Keck | Kp | -49, 217 |
| 2020 Dec 05 12:28 | Keck | J | -49, 204 |
| 2020 Dec 05 12:30 | Keck | H | -49, 200 |
| 2020 Dec 05 12:35 | Keck | Kp | -49, 193 |
| 2020 Dec 05 12:43 | Keck | J | -49, 182 |
| 2020 Dec 05 12:45 | Keck | H | -49, 179 |
| 2020 Dec 05 12:47 | Keck | Kp | -49, 176 |
| 2020 Dec 05 12:55 | Keck | J | -49, 165 |
| 2020 Dec 05 12:57 | Keck | H | -49, 162 |
| 2020 Dec 05 12:59 | Keck | Kp | -49, 159 |
| 2018 Apr 24 09:00 | VLT/SPHERE | N_R | +73, 26 |
| 2018 Apr 28 07:43 | VLT/SPHERE | N_R | +74, 180 |
| 2018 May 04 06:02 | VLT/SPHERE | N_R | +75, 210 |
| 2018 May 05 01:51 | VLT/SPHERE | N_R | +75, 209 |
| 2018 Jun 04 00:07 | VLT/SPHERE | N_R | +80, 254 |
| 2019 Jul 28 09:13 | VLT/SPHERE | N_R | -10, 307 |
| 2019 Jul 30 06:19 | VLT/SPHERE | N_R | -10, 37 |
| 2019 Jul 30 08:03 | VLT/SPHERE | N_R | -10, 248 |
| 2019 Aug 03 04:48 | VLT/SPHERE | N_R | -9, 210 |
| 2019 Aug 06 04:28 | VLT/SPHERE | N_R | -8, 182 |

2015 Keck data are from Drummond et al. [2018] and Shepard et al. [2017]. 2020 Keck data are sums of 3 images, each ~7s exposure, acquired over an interval of 45-55s.  VLT data are from Viikinkoski et al. [2018], Ferrais et al. [2020] (https://observations.lam.fr/astero/16Psyche/). Times are UTC midpoints of each observation or sums of observations. The pixel scale for Keck images was 0.0094" per pixel; VLT was 0.0036" per pixel. Lat/Lon is the model sub-observer latitude and longitude at the time of the observation (light-time corrected).

**Figure 4. VLT observations of** Psyche along with the shape model fits. VLT observations in 2018 are in the left 2 columns (northern hemisphere) and 2019 observations are in the right two columns (equatorial). In each pair of columns, observations are on the left and model fits are to the immediate right. Model features are noted by arrows and the first letters of feature names (Table 6). Observations are chronological from top-to-bottom. Alpha, Bravo, Charlie, and the major axis (0°) are indicated on the first of the 2018 observations for reference. Sub-observer longitudes are above the 2019 observations (latitudes -10°). Details are in Table 4. Image scales are 6.0 km/pixel (2018), 4.5 km/pixel (2019). One of the optical "bright spots" reported by Ferrais et al. is visible in the center of the *lon*=307° image.

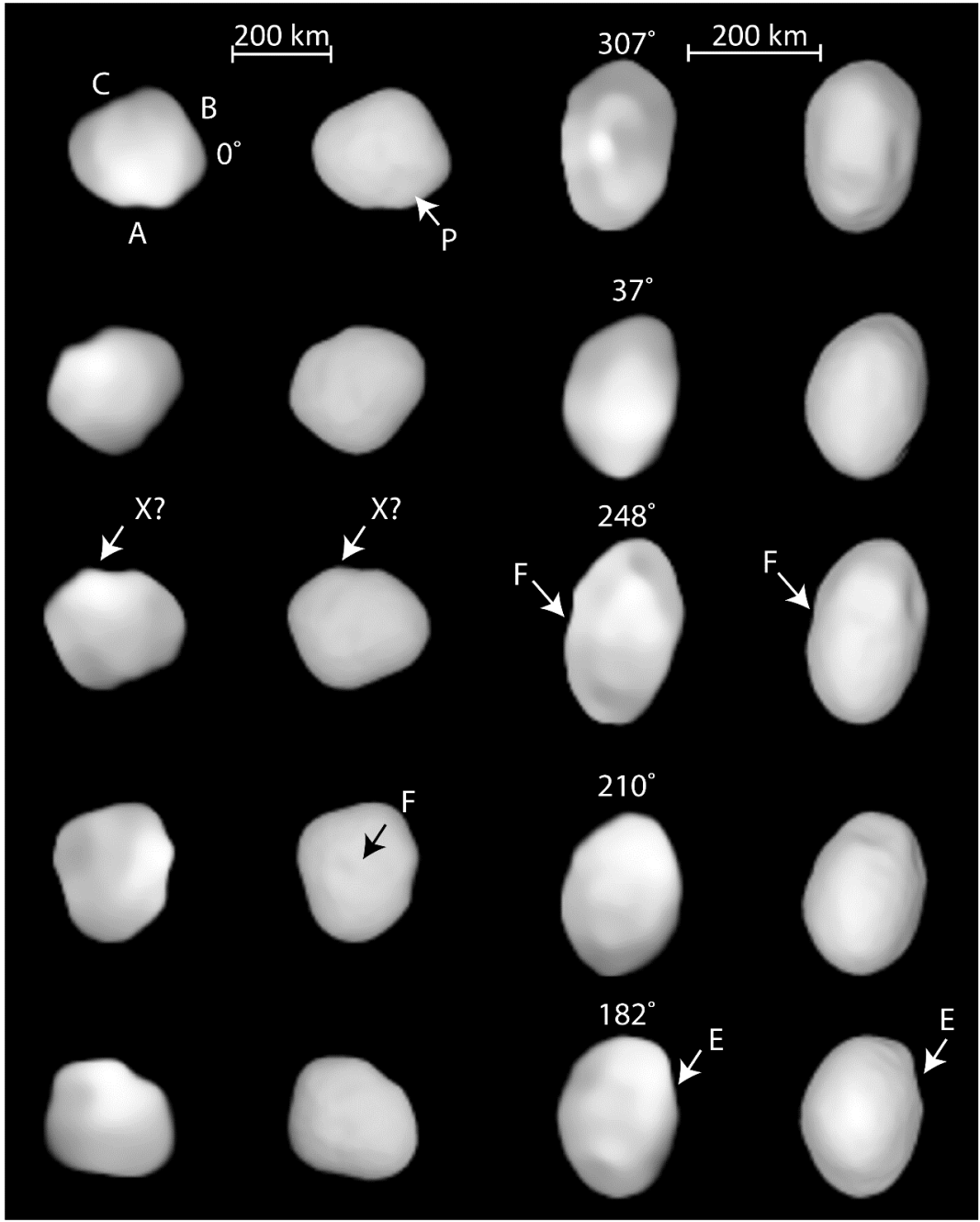

**Figure 5.** Keck AO images of Psyche and the associated shape model fits. Images are in column pairs: observations are on the left and model fits are to their immediate right. At the top of each column pair, the observation date and filter is noted. Within each column, observations are sorted chronologically from top to bottom. 2015 observations are described in Shepard et al. [2017] and Drummond et al. [2018] and were used in the fit. 2020 model images are post-fit predictions. Dates, times, and sub-observer aspects are in Table 4. Image scale is 12 km/pixel.

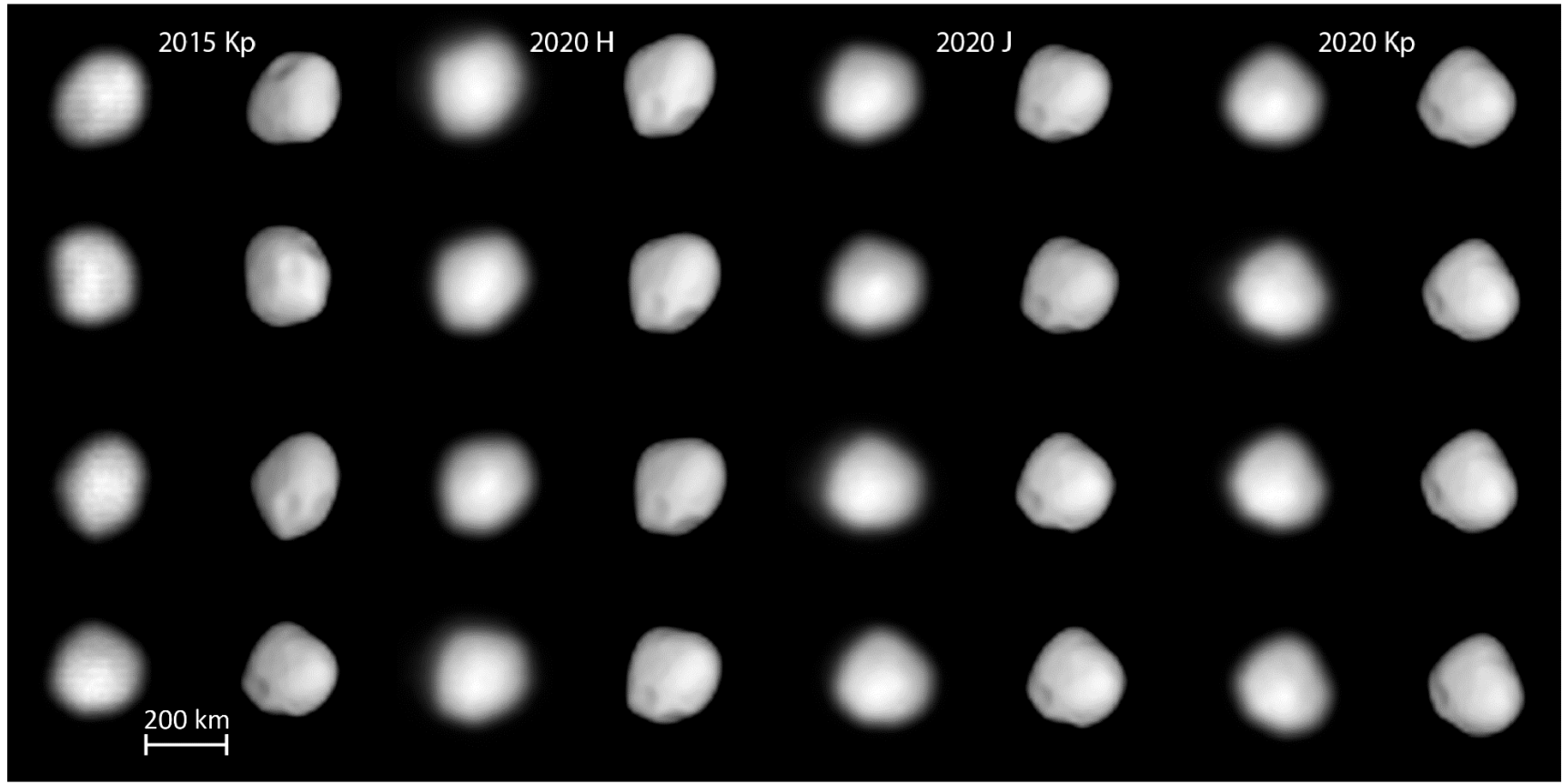

## *2.4 2019 and 2010 Occultations*

On 2019 October 24, Psyche occulted a 10[th] magnitude star and a well-organized campaign led by members of the International Occultation and Timing Association (IOTA) generated 15 evenly spaced and highly accurate chords[7]. This occultation is similar in its equatorial aspect to occultations in 2009 and 2014, but is superior to those in both the quantity and quality of chords. In addition to tightly constraining the c-axis, the 2019 occultation was effectively broadside (shape model sub-observer longitude 95°) and helped to constrain the major (a-) axis. We also include a post-fit comparison to a previous occultation on 21 August 2010[8] which was at a southern aspect (-53°, 350°) and had a dozen chords that spanned the object (**Figure 6**).

**Figure 6.** Comparison of Psyche shape model and the 2019 October 24 (left, aspect -5°, 95°) and 2010 Aug 21 (right, aspect -53°, 350° and not used in the fit) occultations. The misfit at the bottom of the 2010 occultation is likely an offset timing error. The regions labeled "Foxtrot" and "Xray" are described in the text. North is up in both figures.

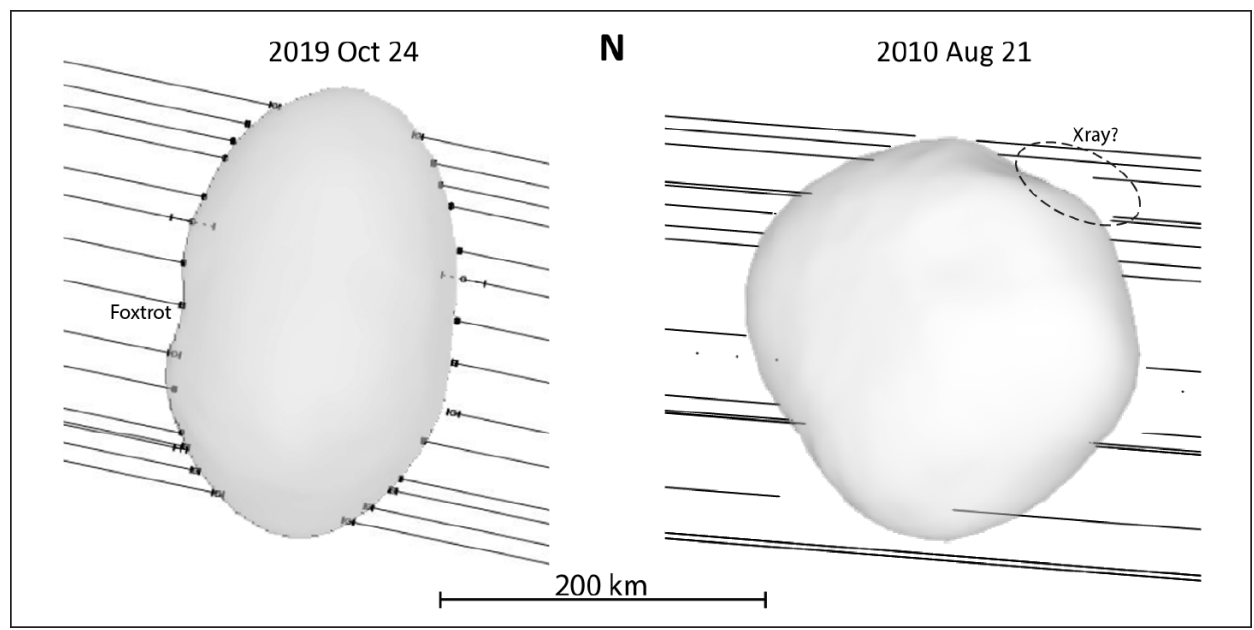

*2.5 Methods*

We utilized the SHAPE modeling software and strategies described by Magri et al. [2007b] and Shepard et al. [2017]. This software simulates the radar image or echo power spectrum for a model shape and spin state and compares it to the images or spectrum taken at the same time. It is also capable of generating synthetic plane-of-sky images that can be compared with images from optical systems. Given an initial shape input, the software uses a gradient search iterative process and adjusts the size, aspect ratios, scattering law(s), spin rate, and pole direction to minimize chi-square, the root-mean-square of the differences between the observations and model.

When beginning the modeling process, we start with ellipsoidal models of many different sizes, aspect ratios, and spin parameters, (i.e., a grid of parameters) and find those combinations that best fit the observations. This reduces the chance of falling into a local minimum solution space.

Once the approximate size, aspect ratios, and spin parameters are better constrained, the ellipsoids are converted into more flexible spherical harmonic shape models to flesh out the gross deviations from an ellipsoid. When these no longer improve, the best models are converted into faceted vertex models for additional fine tuning. In this case, we also explored the results when starting with the previous shape models of Shepard et al. [2017] and the Ferrais et al. [2020].

With spherical harmonic and vertex models, penalty weights are used to enhance or minimize features on the model, including surface roughness or concavities. Different models may be indistinguishable to the chi-square formalism, so the final model is chosen based on its chi-square *and* apparent visual goodness of fit to individual features. We also use post-fit comparisons to other datasets that could not be directly used, such as occultations, as a further check on final size and shape.

For our revised shape model, we included the following datasets: from Arecibo, the 2015 delay-Doppler images; from Keck, the four 2015 adaptive optics images used by Shepard et al., [2017]; from VLT, ten deconvolved adaptive optics images selected from the 2018 and 2019 campaign [Vernazza et al. 2018; Viinkinkoski et al. 2018; Ferrais et al. 2020]; from ALMA 22 plane-of-sky observations acquired in 2019; and the 2019 occultation.

For the occultation, the coverage was extensive enough to generate a plane-of-sky silhouette that was used as a plane-of-sky observation (this was not possible with the other occultations). This required artificially moving the Sun for that dataset to put it at opposition and remove any (cast) shadows in the model fit. Similarly, the ALMA images are due to emission (not reflected light) and the position of the Sun for those data was also altered to simulate an opposition view. In both cases, the primary goal was to constrain Psyche's shape outline and dimensions. For all data sets, we allowed radar and optical scattering models to float so that the shape was the primary quantity of interest to the fit.

Psyche is a rapid rotator (period ~4.2h) and some of our data sets were subject to rotational smearing. For example, each radar imaging run integrated 28 minutes of data, corresponding to ~40$^{\circ}$ of rotation. To minimize the effects of smearing in the final model, we adopted two strategies. First, we had the SHAPE software break up the radar integration time into three snapshots of the model (~13$^{\circ}$ of rotation between them), synthesize the resulting radar images at each time, and combine them to effectively model the smearing due to rotation during the total integration time. Second, the AO, ALMA, and occultation datasets are snap-shots of Psyche with integration times of a few seconds to a few minutes and little or no smearing. By weighting these datasets more (~75%) than the reduced-smear delay-Doppler images (~25%) in the final model, we effectively reduced any smear to scales on the order of our image resolution.

## 2.6 Results

Our final model has dimensions ($a$ x $b$ x $c$) of 278 (−4/+8 km) x 238 (−4/+6 km) x 171 km (−1/+5 km) and an effective spherical diameter of $D_{eff}$ = 222 −1/+4 km. Our uncertainties reflect our best estimates of the possible range of sizes. They are asymmetric because, while a range of sizes were found to be compatible with the data, our best model fell in the lower half of that range. Our range of uncertainties along the a- and b-axes are based on the highest resolution AO and ALMA images; our uncertainties along the c-axis are tighter because of the exceptional coverage of the 2019 occultation.

Our model is consistent in size and shape with those of Drummond et al. [2018], Viikinkoski et al. [2018] and Ferrais et al. [2020]. It is also consistent in size along the major and intermediate axes and in the general shape of the Shepard et al. [2017] radar-derived model, but it is approximately 10% shorter than that model along the *c-axis*.

Our model's spin axis is (ecliptic) (36°, −8°) ± 2° and is consistent with all recently published models. Our uncertainty is based primarily on the 2019 occultation; even a 1° shift in the pole visibly changed the apparent orientation of our model with respect to the occultation chords.

The properties of the final model are listed in **Table 5**. Views of the model fit to the radar, ALMA, VLT, Keck, and occultation data are shown in **Figures 1 and 3-6**. Comparative views along the principal axes for this model are shown along with those of Shepard et al. [2017] and Viikinkoski et al. [2018]/Ferrais et al. [2020] in **Figure 7.**

Using a mass estimate for Psyche of 22.87 ± 0.70 x $10^{18}$ kg [Baer and Chesley, 2017; Elkins-Tanton et al. 2020], we calculate a bulk density 4000 kg $m^{-3}$. This is consistent with previously published estimates [Elkins-Tanton et al. 2020]. The most recent mass estimate is 22.21 ± 0.78 x $10^{18}$ kg [Siltala and Granvik, 2021] which is within the uncertainties of the Baer and Chesley [2017] value. We therefore adopt a bulk density of 4000 ± 200 kg $m^{-3}$.

**Table 5. Psyche Model Characteristics**

| Parameter | Value |
| --- | --- |
| Maximum dimensions (km) | 278 x 238 x 171 |
| Uncertainties (km) | -4/+8, -4/+6, -1/+5 |
| $D_{eff}$ (km) | 222 -1/+5 km |
| DEEVE (km) | 274 x 234 x 171 |
| Surface area ($km^2$) | 1.62 ± 0.05 x $10^5$ |
| Volume ($km^3$) | 5.75 ± 0.19 x $10^6$ |
| Mean visual albedo, $p_v$ | 0.16 ± 0.01 |
| Sidereal rotation period (h) | 4.195948 ± 0.000001 |
| Ecliptic Pole (λ,β) | (36°, -8°) ± 2° |
| Equatorial Pole (α,δ) | (36.3°, +6.1°) ± 2° |
| $W_0$ (TDB) | 270.4° |

Based on a final model of 1652 vertices and 3300 facets. Uncertainties indicate our best estimates of the possible sizes and are asymmetric for reasons described in the text. $D_{eff}$ is the diameter of a sphere with the same volume as the model. DEEVE is the dynamically equivalent equal-volume ellipsoid, the ellipsoid with the same volume and moments of inertia as the model. Mean visual albedo assumes an absolute magnitude $H$ = 5.90 (JPL Small Body Database). Pole coordinates are given in ecliptic and converted to the equatorial system to comply with the recommended IAU format for spin characteristics. $W_0$ is the location of the prime meridian (major axis) at J2000 Barycentric Dynamical Time (TDB). The Psyche shape file (.obj format) is included in the online supplemental materials.

**Figure 7.** Views of the previous shape models of Shepard et al. [2017], Viikinkoski et al. [2018]/Ferrais et al. [2020] (ADAM), and this work. Numbers at the bottom indicate the body-centered longitudes facing the viewer. North is up for the left and center views. Numbers to the left indicate the longitude of the axis pointing in that direction. The polar views (right column) are aligned so that the model major axis is horizontal. Note the similarity in a/b extent and shape. However, the Shepard et al. model is approximately 10% thicker (c-axis) than the others and shows slight differences in its polar aspect. Data for the Shepard et al. model can be found at: https://echo.jpl.nasa.gov/asteroids/shapes/shapes.html

Data for the Viikinkoski/Ferrais et al. models can be found at https://astro.troja.mff.cuni.cz/projects/damit/asteroid_models/view/1806 [Durech et al. 2010] and https://observations.lam.fr/astero/16Psyche/3DModel/

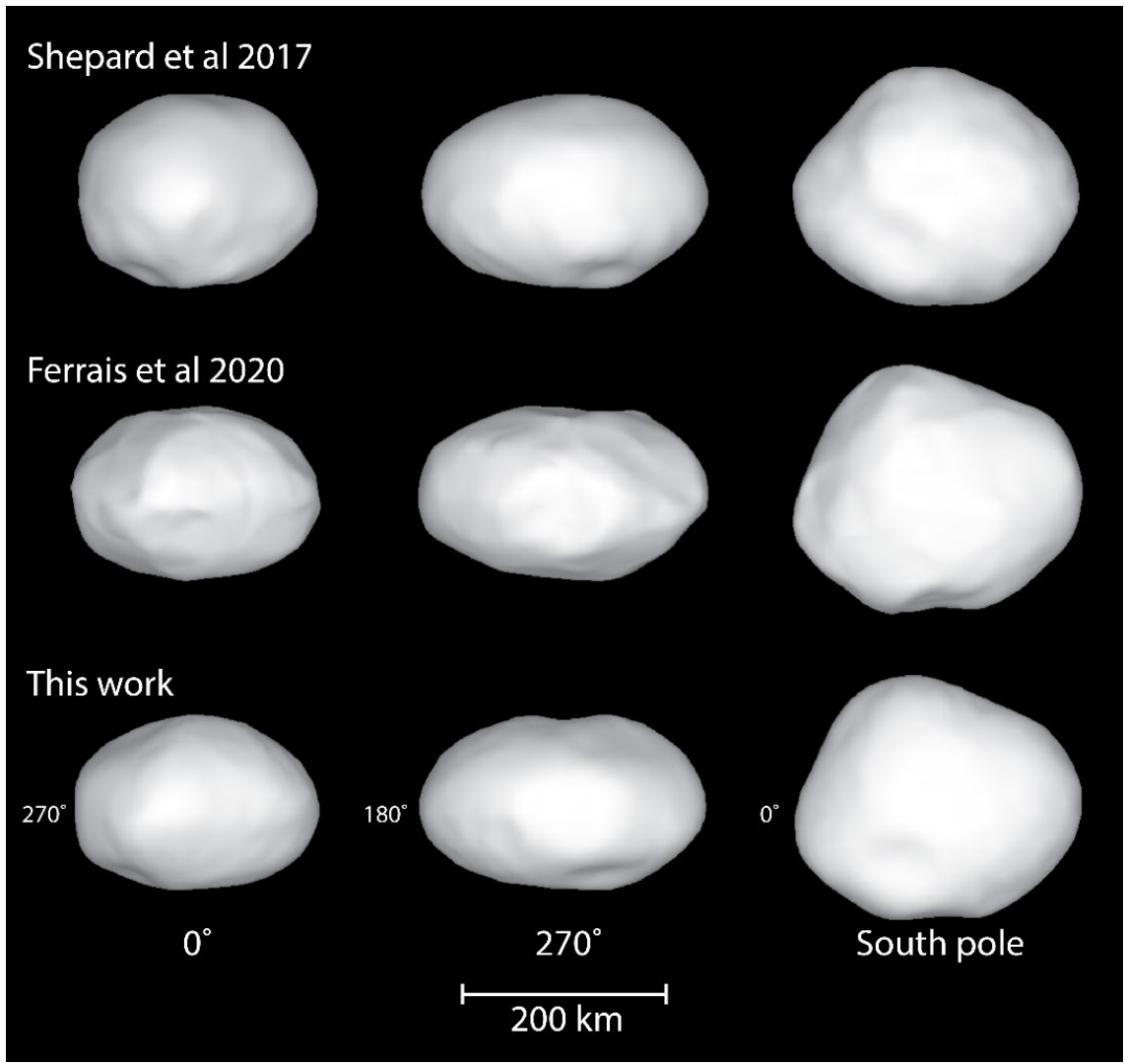

**3. SHAPE AND TOPOGRAPHIC FEATURES**

Here we examine the largest topographical features evident in our model and
those reported by others, and evaluate whether they are likely to be real or
possible artifacts of data processing and inversion. While there are many
subtle depressions evident in our model, it is plausible that some are real,
and equally plausible that some are noise artifacts. For this reason, we will
ignore them unless they are pertinent to the discussion. Where relevant, we
will also note optical albedo features reported by others.

To better organize our description of individual features, we adopt names
from the International Civil Aviation Organization (ICAO) phonetic alphabet
commonly used by radio operators.[9] Details of each feature are listed in **Table
6**. In Figures 1 and 3-6, topographical features are indicated and identified
by their first letter.

We reference Psyche's major features with respect to our shape model body-
centered longitude and latitude, where the +$a$-axis (major) defines 0°
longitude, the +$b$-axis (intermediate) is at 90° longitude, and the +$c$-axis
(minor) aligns with the spin axis in the positive (north)-polar direction.

Ferrais et al. [2020] (their Figure A.6.) noted three equatorial regions of
missing mass relative to an ellipsoidal reference figure, and refer to these
"depressions" as A, B, and C. We follow suit but refer to them as **Alpha**,
**Bravo**, and **Charlie**, respectively. They are visible in our model and are
illustrated in **Figure 8.** Alpha falls around longitude 270° and is the only
one that appears to be a depression while the others are large, flat areas.
Bravo is the missing mass region first identified by Shepard et al., [2017]
and falls between longitudes 340° to 50°. Charlie falls between longitudes
90° and 150°.

Although it is common to treat the shape of asteroids as variations on
ellipsoids, our model of Psyche is also well described (along the major and
intermediate axes) by a rounded rectangular shape with only one corner deviating significantly from this figure (**Figure 8**). The side lengths differ by only ~10%. From this perspective, Bravo and Charlie disappear, and Alpha remains and becomes considerably wider. We will continue to refer to the smaller region as Alpha and the extended area as **West Alpha**.

---

[9]Once spacecraft mapping of Psyche begins in earnest, we propose for consideration
that most feature names be based on those historical figures who have helped to
illuminate the *human* psyche, including those from the broadly defined neurosciences
(psychology, psychiatry, behavioral sciences, etc.) and pioneers in artificial
intelligence. Not only does this convention fit the name, but we feel it holds promise
for exciting public interest and educational outreach.

**Table 6. Topographical and Albedo Features on Psyche**

| Feature Name | Lat, Lon(°) | Notes | Confidence | Reference |
|---|---|---|---|---|
| **Alpha (A)** | 0, 270 | Large depression/crater | Likely | F20, this work |
| **Bravo (B)** | 0, 340->50 | Missing mass region, flat side | Almost certain | S17, V18, F20, this work |
| **Charlie (C)** | 0, 90->150 | Missing mass region, flat side, low optical albedo | Almost certain | V18, F20, this work |
| **Delta (D1)** | -80, 90 | Dynamical depression | Indeterminate | S17 |
| **Eros (D2)** | -65, 260 | Crater (50-75 km) | Almost certain | S17, this work |
| **Foxtrot** | +90 | Crater (50 km) | Likely | This work |
| **Golf** | -45, 15->45 | High radar and optical albedos, possible depression. | -- | This work, V18, F20 |
| **Hotel** | -45, 120->130 | High radar and optical albedo spots, flat topography | -- | F20, this work |
| **India** | -45, 240->300 | Bifurcated radar echoes, elevated radar albedo, high optical albedo | -- | V18, F20, this work |
| **Meroe** | 0->20, 90->120 | Crater (80-100 km), low optical albedo | Indeterminate | V18 |
| **Panthia** | ~40, 300 | Crater (~90 km), high optical and high radar albedos. | Almost certain | V18, this work |
| **Xray** | 0, ~270 | Unknown topography, optically dark | Indeterminate | This work |

Feature names in parentheses are those used by the original reference indicated. Lat/Lon is referenced to this work's shape model; approximate longitudinal extents are indicated by ->. Notes indicate the type of feature suggested and, in some cases, an associated optical or radar albedo. Confidence is a qualitative judgement on whether a topographical feature exists or is an artifact. Dashed lines are shown for features defined originally (or only) by optical or radar albedo. S17 is Shepard et al. [2017]; V18 is Viikinkoski et al. [2018]; F20 is Ferrais et al. [2020].

**Figure 8.** The general shape of Psyche viewed from the South pole. The view on the left shows an ellipsoidal overlay, the major and intermediate axes with longitudes labeled, and the regions referred to as A, B, and C by Ferrais et al. [2020] and here referred to as Alpha, Bravo, and Charlie, respectively (see Table 6). The view on the right shows an alternative rounded rectangular overlay which fits well everywhere except between longitudes 200° to 290°. With this perspective, Bravo and Charlie are not depressions or regions of missing mass, and Alpha becomes considerably larger. The extension of Alpha is referred to as Alpha West or West Alpha.

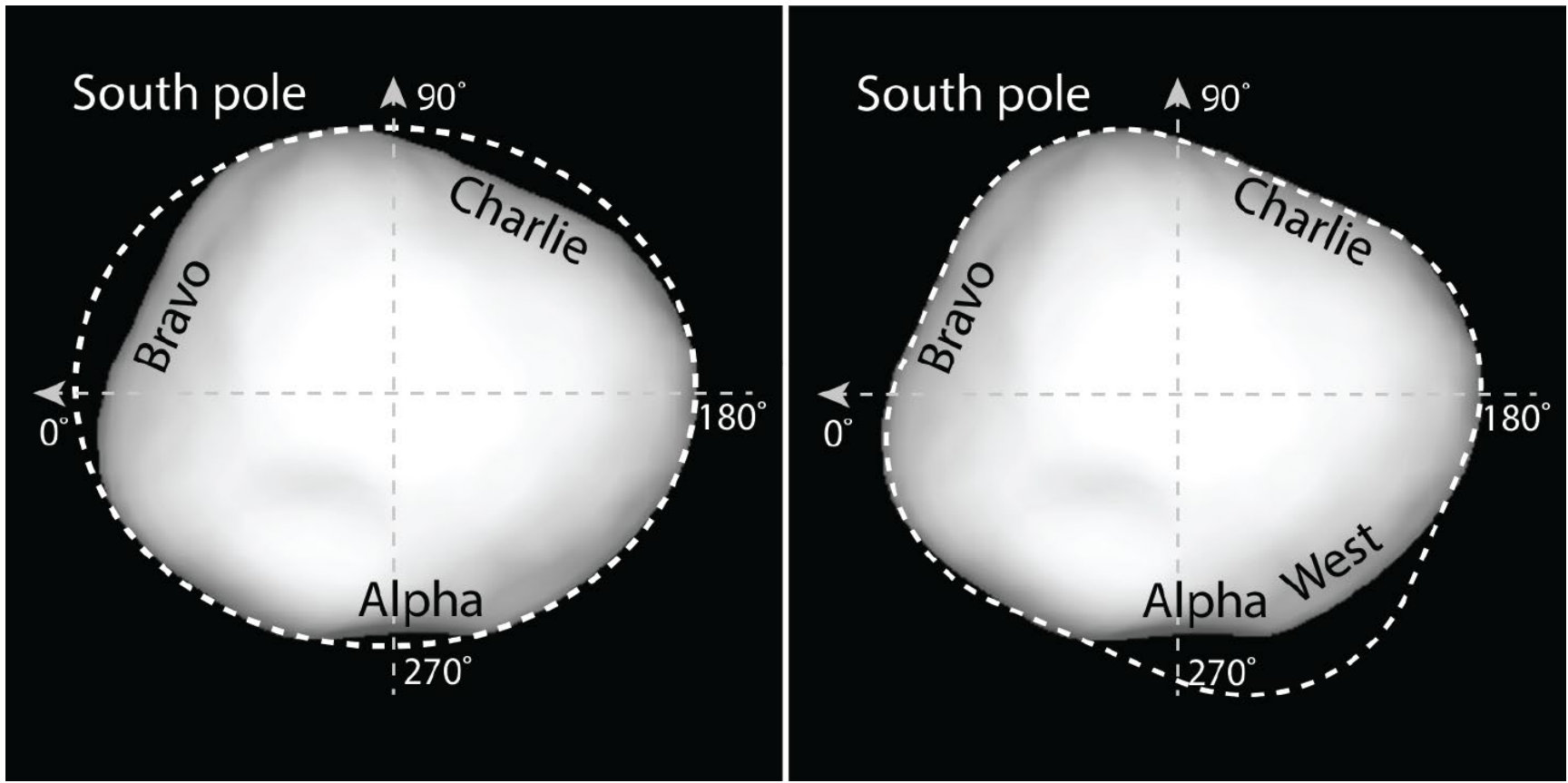

**Figure 9.** Screenshots of Psyche from two animated files (MP4, 18 seconds each) illustrating several features discussed in the text and Table 6 (labels are not present in the animations). The animations cover a full rotation as seen from +20° and -20° latitude. The red peg indicates the +x axis (0°, 0°), and these screenshots are centered on longitude 290°. Both images are oriented with the spin axis up. A Lambertian rendering has been used to emphasize topographic variation.

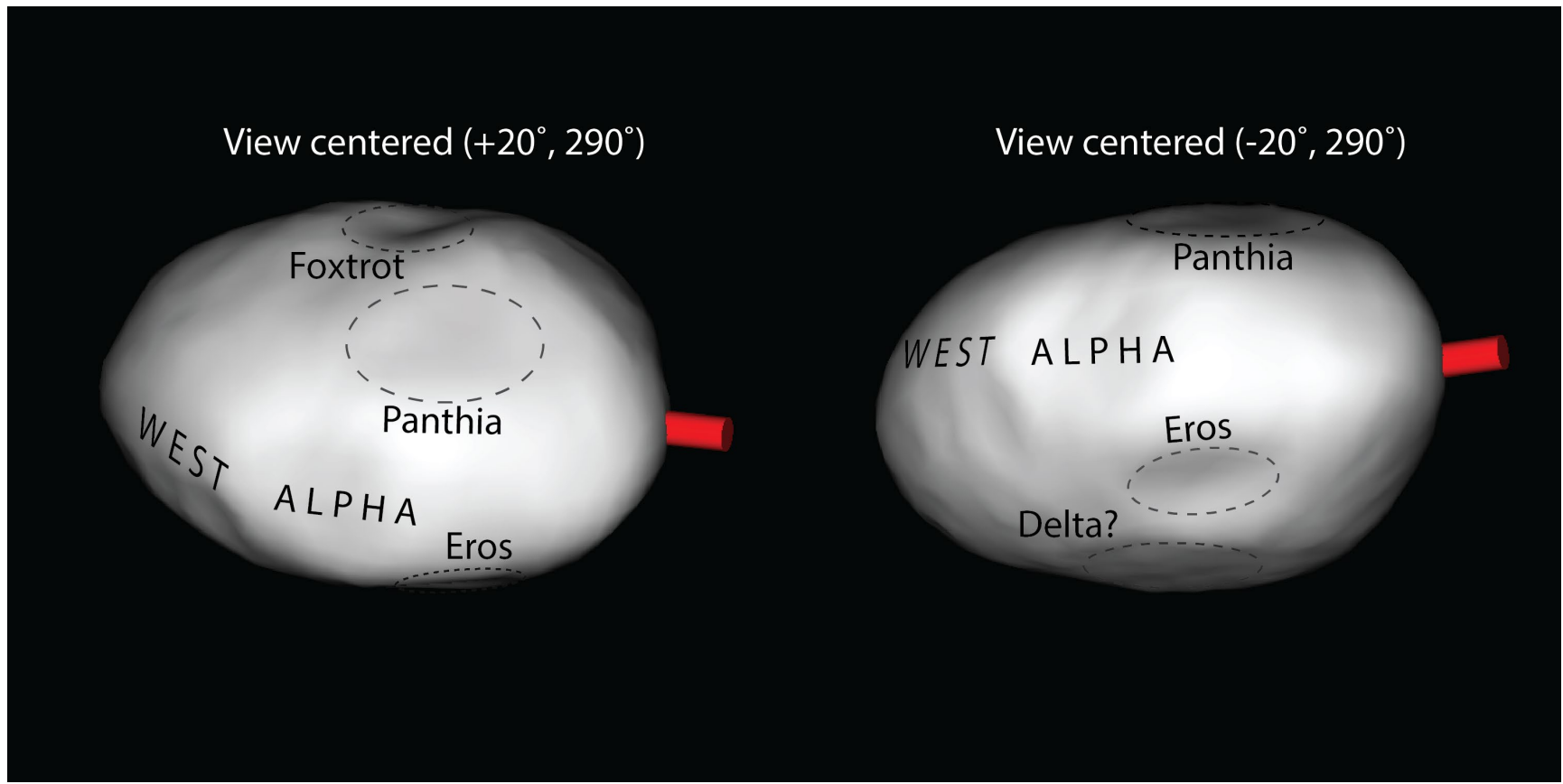

Asteroids 2867 Steins [Jorda et al. 2012], 101955 Bennu [Barnouin, 2019], and 162173 Ryugu [Watanabe et al. 2019] all display shapes with a significant polygonal character, and for the latter two, this is thought to be a consequence of their rubble-pile nature [Michel et al. 2020]. It is unclear whether that is possible for an asteroid the size of Psyche, some two orders of magnitude larger than Bennu or Ryugu. The surprising fit to a rounded rectangle may be a coincidence, or it may reveal something about Psyche's internal structure.

**Figure 9** shows screenshots from two animations (MP4) to aid in visualization. The figure on the left shows the model at a sub-observer latitude and longitude of +20⁰, 290⁰ and the one on the right at -20⁰, 290⁰. The red peg indicates the +x axis (lon 0⁰). Several major features of interest are labeled on the screenshots but not in the animation.

The presence of Alpha is driven primarily by the deconvolved VLT images of the north polar region (**Figure 4**) and is less obvious in the raw data. This flattened-to-concave area should be visible in the ALMA data but is not (**Figure 3**). Instead, the ALMA data suggest there is additional topography at this longitude (270°) that is not readily evident in the VLT or Keck data. The Arecibo observations were not favorably aligned to support either possibility. The ALMA and AO data are complementary in that regions that are dark in visible and near infrared (AO) will be bright in the thermal IR (ALMA), and vice-versa, so perhaps this region is optically dark. We will refer to this possible feature as **Xray**. Its presence is consistent with the shape outlined by the 2010 occultation (**Figure 6**), and the Viikinkoski et al. [2018] albedo map does show a dark region at this longitude. We conclude that the depression Alpha is likely a real feature, though not a certainty, and may include unmodeled topography.

The features referred to as Bravo and Charlie are observed in the radar data, the AO observations from Keck and VLT, and the shape outlined by the 2010 occultation. The evidence for these regions is convincing.

Shepard et al. [2017] noted two *dynamical* depressions in Psyche's southern hemisphere which they labeled D1 and D2. These were regions where the topography, rapid rotation, and estimated gravity of that model conspired to create regions of higher gravitational force, or dynamical depressions. These are locations where fines might preferentially pond and they appeared to correspond with purely topographical depressions. In our model, we find hints but no clear evidence for a topographical depression that coincides with the dynamical depression shown as D1 in that paper (here referred to as **Delta**) (**Figure 9**). As a result, we conclude that a significant topographical depression at the south pole is possible, but indeterminate.

The dynamical depression referred to as D2 in the Shepard et al. [2017] model *is* evident in the topography and is likely to be a large impact crater. It is consistently visible in the delay-Doppler images (**Figure 1**) as a pocket of low SNR pixels, typically 1-3 standard deviations lower than the surroundings. It is also suggested (a side view) in some of the ALMA images (**Figure 3**). We conclude that the evidence for it is convincing and will refer to this depression as **Eros**[10]. In our model, it is centered at ~longitude 290°, ~latitude -65° and appears to be between 50 and 75 km wide and ~4 km deep (**Figure 9**). From some viewing aspects, there are indications that it might be two smaller overlapping depressions.

Our model shows a significant topographical depression at the *north* pole of Psyche and we refer to it as **Foxtrot (Figure 9)**. It was not noted by Viikinkoski et al. [2018] or Ferrais et al. [2020], but signs of it are found in the deviation from ellipsoid figure shown in Ferrais et al. (their Figure A.7). This region was not visible in the data sets used in the Shepard et al. [2017] model. Evidence of its presence can be seen in the mid-line depression seen in the ALMA images (**Figure 3**), a few of the 2019 VLT images (**Figure 4**), and the 2019 occultation (**Figure 6**). We conclude that the existence of Foxtrot is likely, though not certain. If confirmed, measurements on our model suggest it to be ~50 km wide and a few km deep.

Viikinkoski et al. [2018] described two regions that they referred to as **Meroe** and **Panthia.** Meroe is located around longitude 90° just north of the equator and was noted to be significantly darker (optically) than its surroundings. Their model indicated it to be a crater some 80-100 km in size. Our model shows no significant depression at this location and we conclude that a crater here is possible, but indeterminate.

---

[10]This name is reminiscent of Echo, the ICAO alphabetic spelling for E, but is suggested instead as a counterpart to the Psyche crater on 433 Eros.

The region referred to as Panthia is in the northern mid-latitudes from longitudes ~280° to 320° and was found to contain areas much brighter (optically) than the surroundings [Viikinkoski et al. 2018]. Our model also shows this region to contain a relatively wide (~80-90 km) and shallow depression consistent with the Viikinkoski et al. description (**Figures 4** and **9**). Based on its appearance in both models, we conclude that the evidence for Panthia is convincing.

## 4. RADAR PROPERTIES

One of the primary reasons for visiting Psyche is to investigate an object of a type not yet seen – an object that could be the remnant metallic core of an ancient planetesimal. Fortunately, we have 30 calibrated radar echoes that can provide some insight into the potential concentration of metals in the near-surface (**Table 2**).

For calibrated radar echoes, we transmit a circularly polarized continuous wave (CW) signal to the asteroid and measure the echo power in the same (SC) and opposite (OC) senses of polarization. The OC echo is dominated by first surface reflections and is typically the stronger. Reported measurements include the OC radar cross-section, $\sigma_{oc}$, an estimate of the cross-sectional area of a metallic sphere that would produce the same echo power, and the more intuitive OC radar albedo, $\hat{\sigma}_{OC}$, the ratio of the power received from the target relative to that which would be measured from a metallic sphere of the same cross-sectional area at the same distance. To calculate this, we divide the OC radar cross-section by the projected area of the asteroid at the time of its observation.

The ratio of the SC to OC echo is referred to as the circular polarization ratio, or $\mu_c$. It is generally interpreted to be an indicator of near-surface roughness. A low $\mu_c$ (0.0 to ~0.3) is often interpreted to indicate surfaces relatively smooth at the wavelength scale and dominated by first surface reflections. Higher values are associated with surfaces that are thought to be rough at the wavelength scale, display significant volume or multiple scattering, or some combination of these.

In general, radar albedo is thought to be correlated with the near-surface (upper ~meter) bulk density of an object [e.g., Ostro et al. 1985; Shepard et al. 2015]. Given the assumption that larger main-belt asteroids are generally covered by regolith, higher bulk densities imply a greater concentration of iron-nickel in the regolith. Published radar albedos for asteroids range from lows of $\hat{\sigma}_{OC}$ ~0.04 for primitive type (e.g., P) asteroids to highs of ~0.5 for several of the M-class asteroids [Magri et al. 2007a; Shepard et al. 2015]. The relatively common S- and C-class main-belt asteroids have mean radar albedos of $\hat{\sigma}_{OC}$ = 0.14 ± 0.04 and 0.13 ± 0.05, respectively [Magri et al. 2007a].

Using the average of calibrated echo power spectra from 2005, 2015, and 2017, and revised areal cross-sections of Psyche from this shape model, we estimate Psyche's mean OC radar albedo to be $\hat{\sigma}_{OC}$ = 0.34 ± 0.08 although there is considerable variation over the surface. This value is slightly lower than previously reported, $\hat{\sigma}_{OC}$ = 0.37 ± 0.09 [Shepard et al. 2017] because most of the new observations in the northern hemisphere have $\hat{\sigma}_{OC}$ < 0.30. Nevertheless, this value is still more than twice as high as the mean MBA S- and C- class radar albedos. The most reasonable interpretation of this observation is that the near-surface (upper ~meter) of Psyche has a bulk density significantly higher than the typical S- or C-class asteroid, probably caused by an enrichment of iron and nickel [Shepard et al. 2010].

We estimate the mean circular polarization ratio for the 2017 observations to be $\mu_c$ = 0.1 ± 0.1. This value is consistent with a relatively smooth surface at the wavelength scale and echoes dominated by first surface reflections. These measurements are consistent with those measured in 2005 (0.06) and 2015 (0.11) [Shepard et al. 2008; Shepard et al. 2017].

Viikinkoski et al. [2018] generated their shape model from AO observations and lightcurves by simultaneously optimizing for both shape and albedo variegations. They allowed albedo variations of ~20% higher or lower than the mean ($p_v$ = 0.16) and produced a global albedo map. **Figure 10** shows a refinement of that map in Ferrais et al. [2020] (see Figure A.4 in both references). Our shape model corresponds closely to theirs (poles are within 2°, major axes within 3°, see **Figure 7**), so it is a fair representation of where brighter and darker regions can be found on our shape model. Ferrais et al. reported observing three additional bright features in the 2019 AO images that were not indicated on the map. The sizes of those features must have been significant but were not reported, so we have placed stylized spot figures in the locations they report.

**Figure 10** shows Psyche to be optically dichotomous, a feature noted in early lightcurve studies [*e.g.,* Dotto et al. 1992 and references therein]. It is brighter in a band from the equator to the southern mid-latitudes over roughly 180° of longitude ranging from Bravo to West Alpha. A wide belt of bright material between longitudes 300° and 330° also extends upward into the northern mid-latitude region referred to as Panthia. It is brighter than average in the northern mid-latitudes around longitude 120º and the high northern latitudes. The rest of the asteroid is ~average-to-dark except for the two bright spots located just south of the equator at longitudes of ~120° and ~160°.

We have overlain the Figure 10 map with the locations of the sub-radar centers of each calibrated radar CW observation and assigned symbol sizes based on radar albedos. We divided our radar observations into three categories: radar albedos $\hat{\sigma}_{OC} < 0.3$ are a typical background value (although roughly twice as high as a typical S-class asteroid); radar albedos of $\hat{\sigma}_{OC}$ 0.3 – 0.4 are elevated with respect to the background; and radar albedos of $\hat{\sigma}_{OC} > 0.4$ are significantly higher than background.

**Figure 10.** Smoothed and contour-filled optical albedo map of Viikinkoski et al. [2018] and Ferrais et al. [2020] with purported topographical and albedo features labeled (Table 6). Approximate outlines of depressions Eros, Foxtrot, and Panthia are indicated with dashed lines. The sub-radar centers of each calibrated radar echo are shown (Table 2). Optical albedo is shown as multiplicative deviations from the mean ($p_v$ = 0.16 with a range of 0.13-0.19). Three additional optical bright spots noted by Ferrais et al. are indicated with stylized white symbols.

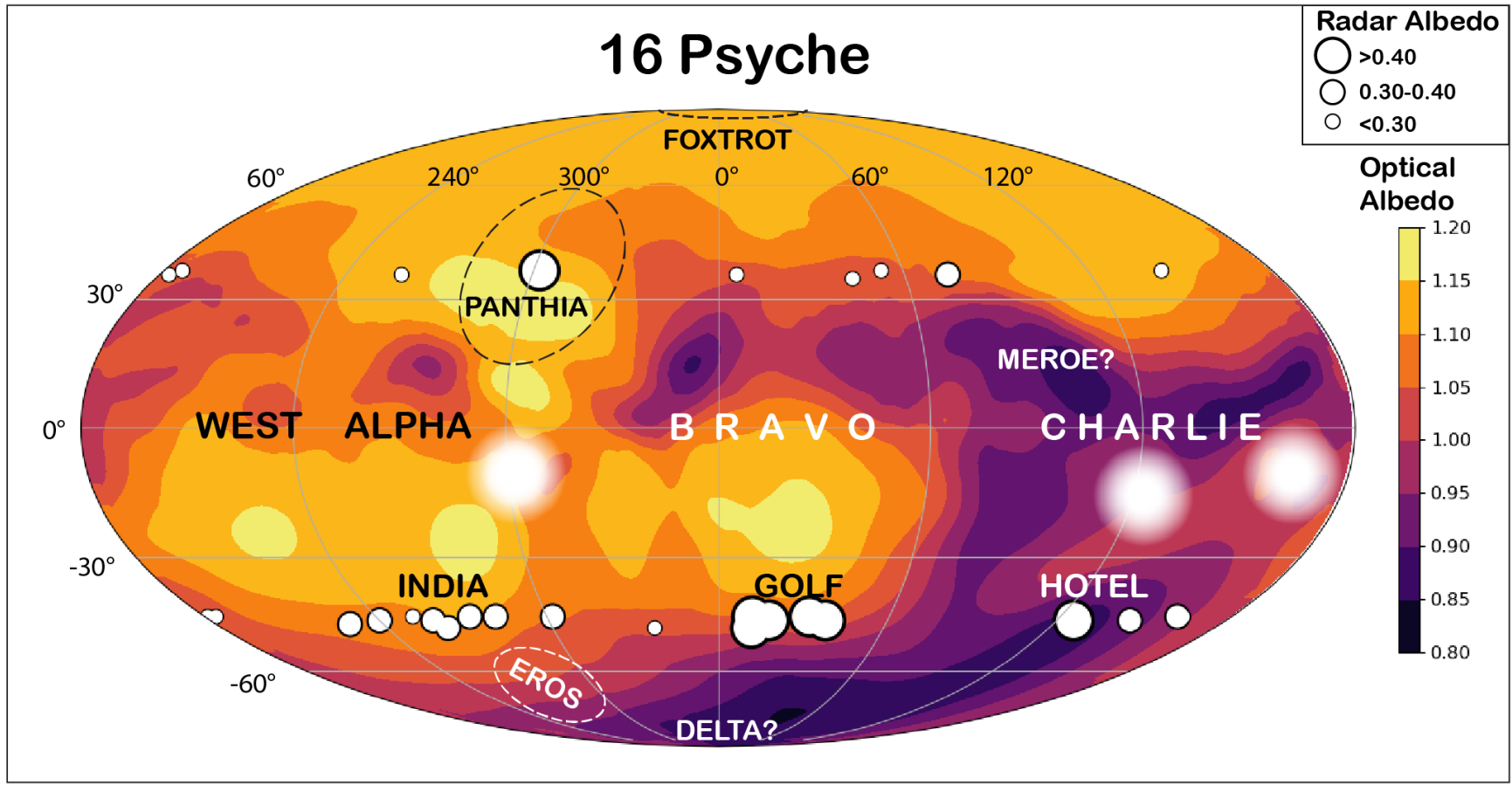

Although the map is coarse, it does reveal at least three areas with significantly higher radar albedos of $\hat{\sigma}_{OC} > 0.4$ (**Table 2**, bolded entries) and at least one region where the radar echoes are consistently elevated ($\hat{\sigma}_{OC}$ of 0.3-0.4) and bifurcated (**Table 2**, bold/italic entries). As a caveat, we note that radar albedo is a whole-disk average; however, the echo will usually be most heavily weighted by the radar properties of features nearest (within ~30°-40°) the sub-radar point.

The first region of high radar albedo is centered on latitude +37° and longitude 303° and corresponds to the topographical depression and optically bright region of **Panthia** (**Table 6**, **Figure 9**, **10**). The regolith may be enriched in metal, or there may be a radar focusing effect because of the depression, or both.

The second region of high radar albedo is centered on latitudes of -45° between longitudes 12° to 50°. Multiple observations on separate dates show it to be a reproducible feature. It falls in an area south of Bravo and we refer to it as **Golf**. There is some evidence for a broad and shallow depression near this region (centered at longitude ~50° south of the equator, **Figure 8**) but it also could be an artifact of our fit. Golf is just south of a region much brighter (optically) than the mean (**Table 6**, **Figure 10**).

The third region of high radar albedo is centered at (-46°, 128°) and is referred to as **Hotel** (**Table 6**, **Figure 10**). Again, multiple observations on separate dates show it to be a reproducible feature. There is no obvious topographical structure here and the region appears generally flat. Overall, this region is optically darker than the mean except for the nearby bright spots noted by Ferrais et al. [2020].

The fourth area of interest is in the southern midlatitudes between longitudes of 230° and 300°, and we refer to it as **India** (**Table 6**, **Figure 10**). It is notable for two reasons: almost all of the echoes in this region have elevated radar albedos and four of them, between longitudes ~260° and ~280°, are bifurcated (**Table 2**, entries with asterisks).

Bifurcated echoes are often associated with a contact binary structure (e.g. Kleopatra, Ostro et al. [2000]; Shepard et al. [2018]), but that is not the case here. Instead, there are likely two or more radar-bright regions separated by a region that is less reflective. As with Golf and Hotel, all the echoes were seen on different dates but fall in the same geographic region, i.e., the behavior is reproducible and is evidence that there is something unusual here.

The bifurcated echoes are centered at the longitude of Alpha, although it is north of our sub-radar points. One echo lobe (on the negative Doppler frequency side) includes reflections from Eros and the bright area and bright albedo spot north of India. The area corresponding to the other lobe (positive Doppler) aligns with the optically bright region just south of West Alpha. There is also some evidence for a broad depression in our model here, just south of the equator and centered at longitude 240°.

There is a possible bifurcated echo in the 2017 observations (northern hemisphere). This echo has an elevated radar albedo like those observed in the south and does fall in a region of modestly higher optical albedo, but there are no obvious structures present.

Based on this admittedly small sample, there appears to be a positive correlation between radar and optical albedos, *i.e.,* the optically brighter regions are more radar reflective and darker regions less so. If confirmed, the most likely reason is that the regolith in the brighter regions has a higher bulk density, and by inference, a higher concentration of metal.

## 5. DISCUSSION AND INTERPRETATION

Here we discuss our two main conclusions and briefly discuss their ramifications for the current models of the formation and evolution of Psyche.

### 5.1 *Psyche is Not Uniformly Radar Bright.*

Shepard et al. [2015] examined radar data from 29 Tholen M-class asteroids, including Psyche, and found that only one-third were "radar-bright", i.e., had mean radar albedos $\hat{\sigma}_{OC} > 0.30$, a property believed to indicate a high metal content in the upper regolith. Psyche and 216 Kleopatra are the largest and most notable of this group. The remaining two-thirds had mean radar albedos $\hat{\sigma}_{OC} \sim 0.25$, including 21 Lutetia ($\hat{\sigma}_{OC}=0.24$). While these values are nearly twice the average value for S- or C-class asteroids, they are inconsistent with a regolith dominated by metal unless it is highly porous [Shepard et al. 2010].

Lutetia was later observed by the Rosetta mission and its appearance and spectrum were found to be more consistent with enstatite chondrite or metal-enriched chondrites like the CH or CB meteorites than with irons [Coradini et al. 2011]. These analogs are among the most widely suggested alternative compositions for the M-class asteroids [e.g., Gaffey, 1976; Hardersen et al. 2011], and all have the moderately elevated metallic concentrations necessary to explain Lutetia's radar albedo [Shepard et al. 2010].

The "background" radar albedo over much of Psyche is $\hat{\sigma}_{OC}$ 0.27 ± 0.03, similar to Lutetia and other non-radar-bright M-class asteroids (**Table 2, Figure 10**). Standing out from this are several local regions that have up to twice the background value. These regions must have even higher bulk densities (e.g., concentrations of metal), and it is because of these regions that Psyche has a significantly higher average radar albedo than Lutetia.

Shepard et al. [2010, 2015] found that nearly all the radar-bright M-class asteroids showed radar scattering behavior like Psyche: rotationally dependent radar albedos, often with bifurcated radar echoes at some aspects indicating distinct radar scattering centers. Except for 216 Kleopatra, none of these objects are known to be contact binaries, and most have the modest lightcurve amplitudes associated with an ellipsoidal shape. A radar image of one of them, 779 Nina, shows an equant object with two separate regions of high radar albedo.

A plausible interpretation of these observations is that Psyche – and possibly any other radar-bright M-class asteroid – has a global silicate regolith with moderately elevated levels of metal, like Lutetia, punctuated with high concentrations of metal in localized regions.

### 5.2 Radar Albedo is Correlated with Optical Albedo

Given the coarse nature of a radar albedo measurement, this conclusion is necessarily tentative, but appears credible. If correct, it provides a clue about the processes operating on Psyche that give rise to localized concentrations of metal. In general terms, this finding seems counter-intuitive because meteoritic iron minerals tend to have dark optical albedos [Cloutis et al. 2010 and references therein]. It may be that the region is brighter because of another mineral phase associated with the process that produced this concentration. Alternatively, there may be a difference in the regolith texture. In general, fine-grain silicates are optically brighter than their coarse-grained equivalents. However, laboratory experiments have found coarse-grained or coherent slabs of meteoritic iron to be optically brighter than fine-grained samples, just the opposite of silicates [Cloutis et al. 2010].

### 5.3 Consequences for Psyche Model Interpretations

Based on the consensus bulk density measured for Psyche, the model of a remnant core is effectively ruled out unless it is a highly porous (~50% microporosity) object [Elkins-Tanton et al., 2020; Siltala and Granvik, 2021]. The unusual shapes of Ryugu and Bennu have been explained as the result of their rubble pile structure [Michel et al. 2020], but it is not known whether metals are strong enough to support a porous structure of Psyche's size [Elkins-Tanton et al. 2020]. The curious fit of our shape model to a rounded rectangle (**Figure 8**) suggests this still may be worth further exploration.

The mechanisms for impact generated regolith on an iron body are still poorly understood, but recent experiments have shown that all the spectral characteristics noted in the past, including the presence of silicates and hydrated phases, are consistent with the formation of glass coatings during hypervelocity impacts of silicates on iron [Libourel et al. 2020]. If composed of pure iron-nickel, Psyche's background radar albedo suggests the upper ~meter of regolith would have to be highly porous (>60% using the model of Shepard et al. [2010]), but this may be consistent with the formation of "foamy" impact melt and carapaces also found in their experiments [Libourel et al. 2020]. Our finding of regional concentrations of metal is also consistent with a hypervelocity impact origin and subsequent evolution. However, of the three highest radar albedo regions, only Panthia has an associated impact structure, although we noted a possible depression at Golf. Hypervelocity impacts might also explain the association of high radar and optical albedos as those experiments found scenarios leading to both darker and brighter optical albedos.

It is conceivable that an impact-gardened silicate-metal regolith [e.g., Davis et al. 1999] would have localized concentrations of metal, but there are several difficulties. If these concentrations are randomly distributed because of some reaccumulation process, it is not readily evident why they are optically brighter. If impacts are invoked, one must explain how they concentrate metal. Impacts could explain the optical brightening as the associated ejecta blanket covers the region in relatively bright silicate fines. However, the rate of space weathering becomes important here [e.g., Clark et al. 2002], for these ejecta fines must stay bright as long as the concentrations are visible.

The interpretation that best appears to fit our observations is that Psyche is a differentiated silicate world, albeit one with elevated metal concentrations consistent with an enstatite or CH/CB chondritic regolith. The radar-bright regions are local ferrovolcanic eruptions of metal as proposed by Johnson et al. [2020], and the surface is optically brighter in these regions because of this process. The iron flow itself might raise the optical albedo because slabs of meteoritic iron have significantly higher albedos than fine grains [Cloutis et al. 2010]. Alternatively, there may be bright secondary materials associated with an eruption, including the volatiles that caused the eruption, compounds derived from them, or materials entrained in the melt from deep in the mantle. Or, like impacts, an eruption may cover a broad area in silicate fines, raising the optical albedo. Here again, the rate of space weathering becomes important.

## 6. SUMMARY AND FUTURE WORK

The earliest interpretation of M-class asteroids like Psyche is that they are the remnant metal cores of ancient protoplanets [Bell et al. 1989]. The consensus after many years of data gathering is that this is unlikely [Elkins-Tanton et al. 2020]. Nevertheless, given the recent experiments of high velocity impacts on metal substrates [Libourel et al. 2020], we find that our results cannot rule it out.

If Psyche is a mixed silicate-metal world, our results are best explained by a differentiated object with a mixed silicate and metal regolith (e.g., enstatite or CH/CB chondritic analogs), peppered with local regions of high metal content from ferrovolcanic activity as proposed by Johnson et al. [2020]. This model provides both a mechanism for concentrating metal in local regions and several possible mechanisms for increasing optical albedos in the vicinity.

Unfortunately, radar observations of Psyche at Arecibo are no longer possible. Goldstone radar can achieve SNRs of ~15/day in 2025, but these are too low for delay-Doppler imaging of Psyche. Prior to the arrival of the Psyche mission in early 2026, Psyche does have favorable oppositions in March 2022 (2.23 AU, similar to the 2017 encounter), May 2023 (2.24 AU), August 2024 (1.70 AU), and December 2025 (1.69 AU, similar to the 2015 encounter) that may provide opportunities for additional AO, ALMA, and other spectral and thermal observations.

**ACKNOWLEDGEMENTS**

We are saddened by the recent loss of the unique Arecibo radar observatory and would like to thank the Arecibo operators and staff for their years of service to the entire astronomical community. Like many others, we advocate for its replacement as an indispensable tool for both characterizing objects within the solar system and for planetary defense against near-Earth objects.

At the time of the 2015 and 2017 radar observations, Arecibo Observatory was operated by SRI International under a cooperative agreement with the National Science Foundation (AST-1100968) in alliance with Ana G. Méndez-Universidad Metropolitana and Universities Space Research Association. The Arecibo planetary radar system was supported by the National Aeronautics and Space Administration under Grant No. NNX12AF24G issued through the Near-Earth Object Observations program. Additional support for radar data analysis and publication is provided by NASA Grant No. 80NSSC19K0523.

This work made use of the JPL Horizons ephemeris service and NASA's Astrophysics Data System.

This paper makes use of the following ALMA data: ADS/JAO.ALMA\#2018.1.01271.S. ALMA is a partnership of ESO (representing its member states), NSF (USA) and NINS (Japan), together with NRC (Canada), MOST and ASIAA (Taiwan), and KASI (Republic of Korea), in cooperation with the Republic of Chile. The Joint ALMA Observatory is operated by ESO, AUI/NRAO and NAOJ. The National Radio Astronomy Observatory is a facility of the National Science Foundation operated under cooperative agreement by Associated Universities, Inc.

We thank M. Viikinkoski for providing the optical albedo data used in Figure 10. We also gratefully acknowledge the use of observations made at the ESO Telescopes at the La Silla Paranal Observatory (VLT) under program 199.C-0074 (PI Vernazza). These data are available at: http://observations.lam.fr/astero/

We thank A. Conrad, J. Drummond, W. Merline and gratefully acknowledge the use of observations from the W.M.Keck Observatory which is operated as a scientific partnership among the California Institute of Technology, the University of California and the National Aeronautics and Space

Administration. The Observatory was made possible by the generous financial support of the W. M. Keck Foundation. The authors wish to recognize and acknowledge the very significant cultural role and reverence that the summit of Mauna Kea has always had within the indigenous Hawaiian community.  We are most fortunate to have the opportunity to conduct observations from this mountain.

We thank E. Cloutis for conversations regarding the spectral nature of metallic minerals.

We also are grateful to the following for observing the 2019 Oct 24 stellar occultation by Psyche: D. Palmer, P. Maley, D. Dunham, J. Dunham, T. George, S. Herchak, R. Jones, R. Reaves, D. Stanbridge, W. Thomas, T. Blank, P. Yeargain, R. Wasson.

**REFERENCES**

Abrahams, J.N.H, Nimmo, F. 2019. Ferrovolcanism: Iron volcanism on metallic asteroids. Geophys. Res. Lett. 46, 5055-5064. doi: 10.1029/2019GL082542

Asphaug, E., Agnor, C.B., Williams, Q. 2006. Hit-and-run planetary collisions. Nature 439, 155-160. Doi:10.1038/nature04311.

Baer, J., Chesley, S.R. 2017. Simultaneous mass determination for gravitationally coupled asteroids. Astron. J 154, doi: 10.2847/1538-3881/aa7de8.

Bell, J.F., Davis, D.R., Hartmann, W.K., Gaffey, M.J. 1989. Asteroids: The big picture. In: Binzel, R.P, Gehrels, T., Matthews, M.S. (Eds.) Asteroids II, Univ. of Arizona, Tucson. pp. 921-948.

Barnouin, O.S. et al. Shape of (101955) Bennu indicative of a rubble pile with internal stiffness. 2019. Nature Geosci 12, 247-252. Doi: _10.1038/s41561-019-0330-x_

Clark, B.E., Hapke, B., Pieters, C., Britt, D. 2002. Asteroid Space Weathering and Regolith Evolution. In Bottke, Jr., W.F., Cellino, A., Paolicchi, P. and Binzel, R.P. (Eds.) Asteroids III, University of Arizona, Tucson, pp. 585-599.

Cloutis, E.A., Hardersen, P.S., Bish, D.L., Bailey, D.T., Gaffey, M.J., Craig, M.A. 2010. Reflectance spectra of iron meteorites: Implications for spectral identification of their parent bodies. Meteor. Planet. Sci 45, 304-332. Dio:10.1111/j.1945-5100.2010.01033.

Coradini, A. et al. 2011. The surface composition and temperature of 21 Lutetia as observed by Rosetta/VIRTIS. Science 334, 492-494. Doi: 10.1126/science.1204062.

Davis, D.R., Farinella, P., Marzari, F. 1999. The missing Psyche family: Collisionally eroded or never formed? Icarus 137, 140-151. Doi: 10.1006/icar.1998.6037.

de Kleer, K., Cambinoi, S., Shepard, M. 2021. The surface of (16) Psyche from thermal emission and polarization mapping. In review, Planetary Science Journal.

Dollfus, A., Mandeville, J.C., Duseaux, M. 1979. The nature of the M-type asteroids from optical polarimetry. Icarus 37, 124-132. Doi: 10.1016/0019-1035(79)90120-9

Dotto, E., Barucci, M.A., Fulchignoni, M., Di Martino, M., Rotundi, A., Burchi, R., Di Paolantonio, A., 1992. M-type asteroids: rotational properties of 16 objects. Astron. Astrophys. Supp. Series 95, 195-211.

Drummond, J.D. et al. 2018. The triaxial ellipsoid size, density, and rotational pole of asteroid (16) Psyche from Keck and Gemini AO observations 2004-1015. Icarus 205, 174-185. Doi: 10.1016/j.icarus.2018.01.010

Ďurech, J. Sidorin, V., Kaasalainen, M. 2010. DAMIT: a database of asteroid models. Astron. Astrophys. 513, doi: 10.1051/0004-6361/200912693.

Elkins-Tanton L. et al. 2017. Asteroid (16) Psyche: Visiting a metal world. 48th Lunar and Planetary Science Conference, The Woodlands, Texas. LPI Contribution No. 1964, id.1718

Elkins-Tanton, L. et al. 2020. Observations, Meteorites, and Models: A preflight assessment of the composition and formation of 16 Psyche. JGR Planets 125, e2019JE006296, 6. https://doi.org/10.1029/2019JE006296.

Ferrais, M. et al. 2020. Asteroid (16) Psyche's primordial shape: A possible Jacobi ellipsoid. Astron. & Astrophys. 638, L15. Doi: 10.1051/0004-6361/202038100.

Gaffey, M.J. 1976. Spectral reflectance characteristics of the meteorite classes, J. Geophys. Res., 81, 905-920.

Hardersen, P.S., Gaffey, M.J., Abell, P.A. 2005. Near-IR spectral evidence for the presence of iron-poor orthopyroxenes on the surfaces of six M-type asteroids. Icarus 175, 141-158.

Hardersen, P.S., Cloutis, E.A., Reddy, V., Mothe-Diniz, T., Emery, J.P. 2011. The M-/X-asteroid menagerie: Results of an NIR spectral survey of 45 main-belt asteroids. Meteor. Planet. Sci 46, 1910-1938.

Johnson, B.C., Sori, M.M., Evans, A.J. 2020. Ferrovolcanism on metal worlds and the origin of pallasites. Nature Astronomy, 4, 41-44. doi:10.1038/s41550-019-0885-x.

Jorda, L., Lamy, P.L., Gaskell, R.W., Kaasalainen, M, Groussin, O, Besse, S., Faury, G. 2012. Asteroid (2867) Steins: Shape, topography, and global physical properties from OSIRIS observations. Icarus, 221 1089-1100. Doi: 10.1016/j.icarus.2012.07.035

Libourel, G. Nakamura, A.M., Beck, P., Potin, S., Ganino, C., Jacomet, S., Ogawa, R., Hasegawa, S., Michel, P. 2019. Hypervelocity impacts as a source of deceiving surface signatures on iron-rich asteroids. Science Advances 5, eaav3971

Magri, C., Nolan, M.C., Ostro, S.J., Giorgini, S.J. 2007a. A radar survey of main-belt asteroids: Arecibo observations of 55 objects during 1999-2003. Icarus 186, 126-151.

Magri, C., Ostro, S.J., Scheeres, D.J., Nolan, M.C., Giorgini, J.D., Benner, L.A.M., Margot, J.L. 2007b. Radar observations and a physical model of Asteroid 1580 Betulia. Icarus 186, 152-177.

Matter, A., Delbo, M., Carry, B., Ligori, S. 2013. Evidence of a metal-rich surface for the asteroid (16) Psyche from interferometric observations in the thermal infrared. Icarus 226, 419-427.

McMullin, J.P., Waters, B., Schiebel, D., Young, W., Golap, K. 2007. CASA Architecture and Applications. Astronomical Data Analysis Software and Systems XVI ASP Conference Series, Vol. 376, proceedings of the conference held 15-18 October 2006 in Tucson, Arizona, USA. Edited by Richard A. Shaw, Frank Hill and David J. Bell., p.127

Michel, P. et al. 2020. Collisional formation of top-shaped asteroids and implications for the origins of Ryugu and Bennu. Nature Communications. Doi:10.1038/s41467-020-16433-z

Ockert-Bell, M.E., Clark, B.E., Shepard, M.K., Rivkin, A., Binzel, R. Thomas, C.A., DeMeo, F.E., Bus, S.J., Shah, S. 2008. Observations of X/M asteroids across multiple wavelengths. Icarus 195, 206-219.

Ockert-Bell, M.E., Clark, B.E., Shepard, M.K., Isaacs, R.A., Cloutis, E., Fornasier, S., Bus, S.J., 2010. The composition of M-type asteroids: Synthesis of spectroscopic and radar observations. Icarus 210, 674-692.

Ostro, S.J., Campbell, D.B., Shapiro, I.I. 1985. Mainbelt asteroids:  Dual polarization radar observations. Science 229, 442-446.

Ostro, S.J., Hudson, R.S., Nolan, M.C., Margot, J-L., Scheeres, D.J., Campbell, D.B., Magri, C., Giorgini, J.D., Yeomans, D.K. 2000. Radar observations of asteroid 216 Kleopatra. Science 288, 836-839.

Prettyman, T.H. et al. 2012. Elemental mapping by Dawn reveals exogenic H in Vesta's regolith. Science 338, 242-246.

Reddy, V. et al. 2012. Delivery of dark material to Vesta via carbonaceous chondritic impacts. Icarus 221, 544-559.

Sanchez, J.A. et al. 2017. Detection of rotational spectral variation on the M-type asteroid (16) Psyche. Astron. J. 153:29 (8pp). Doi: 10.3847/1538-3811/153/1/29.

Shepard, M.K. et al. 2008. A radar survey of M- and X-class asteroids. Icarus 195, 184-205.

Shepard, M.K., et al. 2010. A radar survey of M- and X-class asteroids II. Summary and synthesis. Icarus 208, 221-237.

Shepard, M.K. et al. 2015. A radar survey of M- and X-class asteroids. III. Insights into their composition, hydration state, and structure. Icarus 245, 38-55.

Shepard, M.K. et al. 2017. Radar observations and shape model of asteroid 16 Psyche. Icarus 281, 388-403. Doi:10.1016/j.icarus.2016.08.011.

Shepard et al. 2018 A revised shape model of asteroid (216) Kleopatra. Icarus 311, 197-209. Doi: 10.1016/j.icarus.2018.04.002

Siltala, L., Granvik, M. 2021. Mass and density of asteroid (16) Psyche. Astrophys. J. Lett. 909. Doi:10.3847/2041-8213/abe948

Takir, D., Reddy, V, Sanchez, J., Shepard, M., Emery, J. 2017. Detection of water and hydroxyl on M-type asteroid (16) Psyche. Astron.J. 153:31 (6pp) doi:10.3847/1538-3881/153/1/31.

Tholen, D. 1984. Asteroid taxonomy from cluster analysis of photometry. PhD thesis, Univ. of Arizona, Tucson. 150 pp.

Viikinkoski M. et al. 2018. (16) Psyche: A mesosiderite-like asteroid? Astron. Astrophys. 619 (10 pp). Doi: 10.1051/0004-6361/201834091.

Watanabe, S. et al. 2019. High porosity nature of the top-shape C-type asteroid 162173 Ryugu as observed by Hyabusa2. 50[th] LPSC abstract, LPI (2132): 1265.